\documentclass[%
 reprint,
 amsmath,amssymb,
 aps,
 prx,
 floatfix,
]{revtex4-2}

\usepackage{graphicx}   % Include figure files
\usepackage{dcolumn}    % Align table columns on decimal point
\usepackage{xcolor}
\usepackage{bm}         % bold math
\usepackage{hyperref}   % hypertext capabilities
\usepackage[normalem]{ulem} % for \sout{} strikeout
\graphicspath{{pino_v6/}{pino_figures/}}

\newcommand{\pp}{\partial}
\newcommand{\R}{\mathcal{R}}
\newcommand{\Lloss}{\mathcal{L}}
\newcommand{\OmegaT}{\Omega_T}
\newcommand{\br}{\mathbf{r}}
\newcommand{\bD}{\mathbf{D}}
\newcommand{\avg}[2]{\left\langle #1 \right\rangle_{#2}}
\newcommand{\norm}[1]{\left\lVert #1 \right\rVert}
\newcommand{\bn}{\bm n}
\newcommand{\btau}{\bm\tau}
\newcommand{\bv}{\bm v}
\newcommand{\bB}{\bm B}
\newcommand{\bJ}{\bm J}

\newcommand{\grad}{\nabla}
\newcommand{\Lap}{\nabla_\perp^2}
\newcommand{\bzero}{\bm 0}

\begin{document}

\title{Physics-Informed Neural Operator Surrogates for Two-Dimensional
Magnetohydrodynamic Reconnection}

\author{Kushaal Kumar Pothula}
\email{kushaalkp@iitbhilai.ac.in}
\affiliation{Indian Institute of Technology Bhilai,
Chhattisgarh 491002, India}

\author{Arunav Kumar}
\affiliation{Plasma Science and Fusion Center,
Massachusetts Institute of Technology,
Cambridge, Massachusetts 02139, USA}

\date{\today}

\begin{abstract}

Direct numerical simulation of magnetic reconnection is limited by the scale
separation of resistive magnetohydrodynamics. At large Lundquist numbers $S$
the current layer thins as $S^{-1/2}$, forcing fine grids and short time
steps that make parameter scans prohibitively expensive. Deep learning neural operators
offer an alternative by learning the map between function spaces rather than
individual solutions, so that a single trained model returns the state for any
parameter and time at inference cost. We have developed a Fourier Neural Operator (FNO) based Physics-Informed Neural
Operator (PINO) surrogate for two-dimensional compressible, viscous, resistive reconnection in
a wall bounded domain. We condition on the initial state, the Lundquist number,
and a continuous query time. Predicting the magnetic flux function makes
$\grad\!\cdot\!\bB=0$ exact, direct time queries remove autoregressive error
accumulation, and parity-aware spectral differentiation matches the wall
treatment of the reference solver. Trained across
$10^{3}\le S\le2\times10^{5}$ with two values withheld, the surrogate
recovers density, pressure, and guide field to better than $0.2\%$, velocity
to $3.2$ to $3.9\%$, the flux function to $1.0\%$, and the spectral current
density to $7.5\%$. It resolves the thin, narrow current sheet, Alfv\'enic jets, and
spectral resolution beyond mode cutoff. Our surrogate reproduces the
layer-averaged reconnection rate to within $3\%$ through
$S=1.1\times10^{4}$, recovering a reconnection time exponent of $0.497$
consistent with Sweet-Parker scaling. When queried zero-shot on the $2049^{2}$
solver mesh, a full trajectory run is on the order of seconds on one GPU against
several hours for the DNS, over two orders of magnitude faster, establishing
learned operators as a practical route to surveying reconnection across
parameter space.
\end{abstract}

\maketitle

% ======================================================================

\section{Introduction}\label{sec:intro}

Magnetic reconnection occurs when ideal flux freezing breaks down within a localized non-ideal layer. In resistive magnetohydrodynamics, the macroscopic dynamics are Alfv\'enic, the magnetic Reynolds number is large, and non-ideal evolution is confined to a current sheet that is thin relative to the system size. The relevant asymptotic parameter is the Lundquist number $S=LV_A/\eta$, where $L$ is a macroscopic length, $V_A$ is the Alfv\'en speed, and $\eta$ is the magnetic diffusivity. For $S\gg1$, the ideal region stores and transports magnetic flux, while the current sheet controls changes in magnetic topology, energy release, and the reconnection electric field. Resolving these separated scales remains a central challenge in reconnection theory and simulation \cite{sweet1958,parker1957,petschek1964,biskamp2000,kulsrud2005,yamada2010,zweibel2009,ji2011}.

This scale separation appears throughout plasma physics. In space and astrophysical plasmas, reconnection drives solar flares, coronal mass ejections, magnetotail substorms, accretion-disk dynamics, and relativistic outflows \cite{dungey1953,dungey1961,giovanelli1946,priest2000,shibata2011,birn2001,burch2016}. In magnetic-confinement fusion, it underlies tearing modes, sawtooth crashes, disruptions, and edge-relaxation events \cite{furth1963,rutherford1973,kadomtsev1975,wesson1990,hender2007,helander2014}. Although these systems differ in collisionality, geometry, and kinetic scales, any useful model must resolve or accurately represent the current layer that sets the reconnection rate.

In the classical Sweet-Parker ordering, a steady two-dimensional resistive sheet has aspect ratio $\delta_{\rm SP}/L\sim S^{-1/2}$ and a normalized reconnection rate of the same order. This result follows from mass conservation, Alfv\'enic outflow, and Ohmic diffusion across the layer \cite{sweet1958,parker1957,biskamp1986}. Although internally consistent within resistive MHD, Sweet-Parker reconnection is too slow to explain many high-$S$ plasmas. The Petschek model gives a faster rate through a short diffusion region bounded by standing slow shocks, but uniform resistive MHD generally cannot sustain the required localized diffusion region \cite{petschek1964,biskamp1986,kulsrud2005}. The high-$S$ resistive problem is therefore singular even before kinetic effects are included.

Tearing and plasmoid formation make this singular behavior explicit. A long Sweet-Parker sheet becomes unstable above a critical Lundquist number of order $10^4$ \cite{loureiro2007,bhattacharjee2009,samtaney2009}. Linear theory predicts that the fastest growth rate scales as $S^{1/4}$ and the plasmoid number as $S^{3/8}$ \cite{loureiro2007,samtaney2009}. Nonlinear simulations produce secondary islands and current sheets, with reconnection rates that can depend only weakly on $S$ over broad regimes \cite{huang2010,uzdensky2010,ni2010,loureiro2012,huang2013,huang2016}. Theories of ideal tearing and dynamically thinning sheets further show that onset depends on the sheet aspect ratio, formation history, and perturbation amplitude, not just on $S$ \cite{pucci2014,tenerani2015,comisso2016,tenerani2016,huang2017}. The evolving layer, rather than the smooth global flux function alone, therefore controls the dynamics.

%More complete reconnection models include Hall physics, pressure-tensor effects, collisionless layers, guide-field dynamics, three-dimensional structure, and turbulence \cite{cassak2005,daughton2009,daughton2011,zenitani2011,lazarian1999,kowal2009,eyink2011,lazarian2012,lapenta2012}. These effects are essential in many plasmas but are outside the scope of this study. We instead use two-dimensional, compressible, viscous, resistive MHD as a controlled benchmark. This isolates the first question any surrogate must answer: can it recover the thin current layer and its derivatives, or only the smoother bulk fields?

Direct numerical simulation (DNS) provides the reference solution, but its cost follows directly from this ordering. At high $S$, the sheet width scales as $S^{-1/2}$, while the out-of-plane current is $J_z=\nabla_\perp^2\psi$. A small error in the flux function can therefore produce a much larger error in the current. This matters directly because the reconnecting electric field in resistive MHD contains $\eta J_z$. The grid must resolve the current layer, the time step must satisfy Alfv\'enic and diffusive stability limits, and parameter scans over $S$, sheet thickness, viscosity, guide field, and boundary conditions quickly become expensive \cite{toth2000,stone2008,mignone2007,sovinec2004,jardin2010}.

Neural operators are attractive surrogates because they learn maps between functions rather than individual solutions. A Fourier neural operator can represent the parameter dependent map from initial conditions and physical parameters to later plasma states \cite{li2020,li2021,lu2021,kovachki2021,goswami2022}. Physics-informed neural operators add governing equation residuals to the supervised loss, which can help when the equations are known but high-fidelity data are expensive \cite{raissi2019,karniadakis2021,wang2021,li2021pino,goswami2022}. 
%Recent plasma applications have reported good accuracy for bulk fields in incompressible MHD, tokamak surrogate models, camera-based plasma evolution, and edge-plasma simulations \cite{pathak2018,kochkov2021,stachenfeld2021,rosofsky2023,gopakumar2024,kim2024,carey2025,duarte2025}. Whether these models remain reliable for derivative-dominated reconnection diagnostics is still unclear.

We hence develop a Physics-Informed Neural Operator for two-dimensional, compressible, viscous, resistive MHD reconnection in a wall-bounded domain. The model is conditioned on the initial state, Lundquist number, and query time. We represent the in-plane magnetic field through a flux function, which satisfies $\nabla\cdot\mathbf{B}=0$ by construction. We validate the surrogate against DNS across a range of Lundquist numbers and report errors in the primitive variables, magnetic field, current density, reconnection rate, conservation properties, and resolution transfer. 

Section~\ref{sec:setup} defines the physical and numerical problem. Section~\ref{sec:method} describes the neural operator, training objective, and reconnection diagnostics. Section~\ref{sec:results} evaluates the surrogate across the Lundquist-number sweep and identifies where it succeeds and fails. Section~\ref{sec:conclusions} summarizes the implications for learned reconnection models.

\section{Problem setup and governing equations}\label{sec:setup}

We study reconnection in a two-dimensional reversed-field current sheet. The problem is formulated as an initial boundary value problem for the compressible, viscous, resistive magnetohydrodynamic (MHD) equations \cite{huang2016}. The configuration consists of a single planar sheet centered on the neutral line $y=0$, embedded in a uniform guide field and bounded by perfectly conducting walls. The plasma starts from rest, developing the layer from a laminar Sweet-Parker state. Section~\ref{sec:data} describes the reference direct numerical simulation (DNS) used to evaluate the surrogate.

\subsection{Governing equations}\label{sec:governing}

The plasma state is described by the mass density $\rho$, momentum
$\mathbf m=(m_x,m_y,m_z)=\rho\bv$ with velocity $\bv=(u,v,w)$, the thermal
pressure $p$, the in-plane magnetic flux function $\psi$, and the out-of-plane
guide field $g\equiv B_z$. Because $\pp_z=0$, the magnetic field follows from $\psi$
and $g$:
\begin{equation}
  \bB=(B_x,B_y,g),\qquad
  B_x=-\pp_y\psi,\quad B_y=\pp_x\psi,
  \label{eq:Bfromflux}
\end{equation}
Thus, $\mathbf B_\perp=\grad\psi\times\hat{\mathbf z}$ is solenoidal by
construction, and $\grad\!\cdot\bB=0$ holds. The out-of-plane current density and total pressure are
\begin{equation}
  J_z=\pp_xB_y-\pp_yB_x=\Lap\psi,\qquad
  P_T=p+\tfrac12\big(B_x^2+B_y^2+g^2\big),
  \label{eq:JzPT}
\end{equation}
with $\Lap=\pp_x^2+\pp_y^2$. Mass is conserved,
\begin{equation}
  \pp_t\rho+\pp_x m_x+\pp_y m_y=0,
  \label{eq:continuity}
\end{equation}
 and momentum is advanced in conservative flux form. The fluxes include the total pressure and Maxwell stress. Viscosity enters through a Laplacian term
with kinematic viscosity $\nu$:
\begin{equation}
  \begin{aligned}
    \pp_t m_x
    &+\pp_x\!\left(m_x u+P_T-B_x^2\right)
    +\pp_y\!\left(m_x v-B_xB_y\right)=\nu\,\Lap m_x,\\[4pt]
    \pp_t m_y
    &+\pp_x\!\left(m_y u-B_xB_y\right)
    +\pp_y\!\left(m_y v+P_T-B_y^2\right)=\nu\,\Lap m_y,\\[4pt]
    \pp_t m_z
    &+\pp_x\!\left(m_z u-gB_x\right)
    +\pp_y\!\left(m_z v-gB_y\right)=\nu\,\Lap m_z.
  \end{aligned}
  \label{eq:momentum}
\end{equation}
Thermal pressure obeys an adiabatic energy equation, with adiabatic index
$\gamma=5/3$; explicit Ohmic and viscous heating are omitted:
\begin{equation}
  \pp_t p+\pp_x(pu)+\pp_y(pv)+(\gamma-1)\,p\big(\pp_x u+\pp_y v\big)=0.
  \label{eq:pressure}
\end{equation}
The in-plane field evolves through the scalar induction equation for the flux
function:
\begin{equation}
  \pp_t\psi+u\,\pp_x\psi+v\,\pp_y\psi=\eta\,\Lap\psi=\eta J_z,
  \label{eq:induction-psi}
\end{equation}
The guide field obeys the conservative induction equation
\begin{equation}
  \pp_t g+\pp_x\!\left(ug-wB_x\right)+\pp_y\!\left(vg-wB_y\right)=\eta\,\Lap g,
  \label{eq:induction-g}
\end{equation}
with resistivity $\eta$. Equations~\eqref{eq:continuity}-\eqref{eq:induction-g}
close the system for $(\rho,\mathbf m,p,\psi,g)$. The out-of-plane velocity $w$
enters the magnetic evolution only through the guide-field transport term in
Eq.~\eqref{eq:induction-g}.

We use the nondimensional units of Ref.~\cite{huang2016}, setting the reference
length, velocity, density, magnetic field, and pressure to unity
($L_*=V_*=\rho_*=B_*=p_*=1$). The Alfv\'en speed
$V_A=B_*/\sqrt{\rho_*}$ and Alfv\'en time $\tau_A=L_*/V_*$ are therefore also
unity, and time is measured in Alfv\'en times. The magnetic Prandtl number is
fixed at $\mathrm{Pr}_m=\nu/\eta=1$, leaving the Lundquist number
$S=L_*V_A/\eta=1/\eta$ as the only transport parameter. The Sweet-Parker layer
half-width scales as $\delta_{\rm SP}\sim S^{-1/2}$ and sets the smallest
dynamically relevant length that the initial data, boundary treatment, and
surrogate must resolve.

This setup primarily targets the laminar Sweet-Parker regime. Above the
critical Lundquist number $S_c\approx4\times10^4$, a Sweet-Parker sheet becomes
linearly unstable to the plasmoid, or secondary-tearing, instability
\cite{huang2010}. Its subsequent evolution produces multiple current sheets and
a reconnection rate that can become nearly independent of $S$; the outcome also
depends on the perturbation level rather than on $S$ alone \cite{huang2010}.
Because our initial state is unseeded and the integration window is limited to
$T=2$, the instability must grow from round-off noise. The present DNS and
surrogate therefore resolve the subcritical and marginal regimes more fully
than the developed plasmoid regime at $S\gtrsim10^5$. Studies designed for the
latter usually use a reduced, nearly incompressible system, seed the instability
with controlled low-amplitude noise, and characterize the resulting sheets and
plasmoids statistically. Applying the surrogate to that regime will require the
same seeded, ensemble-based treatment which is beyond the scope of our study. 
\subsection{Computational domain and boundary conditions}\label{sec:bc}

The spatial domain is the square $\Omega=[-\tfrac12,\tfrac12]^2$, and the
integration interval is $t\in[0,T]$ with $T=2$, so
$\OmegaT=\Omega\times[0,T]$. All four boundaries are impenetrable, free-slip,
perfectly conducting walls. These conditions impose zero normal flow,
free-slip tangential velocity, a flux-surface condition on $\psi$, and
homogeneous Neumann conditions on $\rho$, $p$, and $g$:
\begin{equation}
  \begin{aligned}
    \text{on } x=\pm\tfrac12:\quad
      & u=0,\quad \pp_x v=\pp_x w=0,\\
      &\psi=0,\quad
        \pp_x\rho=\pp_x p=\pp_x g=0,\\[4pt]
    \text{on } y=\pm\tfrac12:\quad
      & v=0,\quad \pp_y u=\pp_y w=0,\\
      &\psi=0,\quad
        \pp_y\rho=\pp_y p=\pp_y g=0,
  \end{aligned}
  \label{eq:bc}
\end{equation}
These conditions are supplemented by $\pp_{nn}\psi=0$ on $\partial\Omega$,
where $\pp_n$ is the wall-normal derivative. Because $\psi$ is fixed at zero on
each wall, every boundary is a magnetic flux surface with $\bn\cdot\bB=0$, and
no magnetic flux crosses the domain boundary.

\subsection{Spectral differentiation on the bounded domain}\label{sec:specdiff}

The diagnostics in Sec.~\ref{sec:reconny0} and the training objective in
Sec.~\ref{sec:loss} require spatial derivatives. A direct Fourier derivative is
not appropriate because $\Omega$ is wall bounded rather than periodic. If a
field has different values on opposite walls, the discrete transform introduces
a jump at the boundary, and differentiation spreads the resulting Gibbs
oscillations across the domain \cite{gottlieb1997}. We avoid this artifact by using the wall parity
specified in Eq.~\eqref{eq:bc}: homogeneous Neumann fields ($\pp_nf=0$) are
even, while homogeneous Dirichlet fields ($f=0$) are odd. Reflecting a field of
parity $s\in\{+1,-1\}$ across both walls along one axis extends it from
$[-\tfrac12,\tfrac12]$ to $[-1,1]$:
\begin{equation}
  f^{e}(x)=
  \begin{cases}
    s\,f(-1-x), & x\in[-1,-\tfrac12),\\
    f(x), & x\in[-\tfrac12,\tfrac12],\\
    s\,f(1-x), & x\in(\tfrac12,1],
  \end{cases}
  \label{eq:reflect1d}
\end{equation}
The two-dimensional extension applies the same construction along $x$ and $y$
using the axis-dependent parities in Eq.~\eqref{eq:bc}. The fields $\rho$, $p$,
$g$, and each wall-tangential velocity component are even; $\psi$ and the
wall-normal velocity component are odd. The extended field is continuous and
periodic on $[-1,1]^2$, so its Fourier coefficients decay spectrally rather than
algebraically. We compute $\pp_af$ from
$\widehat{\pp_af^{e}}=ik_a\hat f^{e}$ and restrict the inverse transform to
$\Omega$. The same construction gives second and mixed derivatives, including
$\Lap\psi=J_z$. The reference solver uses the same parity rules locally in its
ghost cells (Sec.~\ref{sec:data}); applying them globally here improves the
current sheet curvature estimate from second-order to spectral accuracy.

\subsection{Dataset generation}\label{sec:data}

We initialize the plasma uniformly and at rest:
\begin{equation}
  \rho_0=1,\qquad p_0=1,\qquad g_0=1,\qquad u_0=v_0=w_0=0,
  \label{eq:ic-uniform}
\end{equation}
Thus, the initial flow is zero and the reconnection layer is not kinematically
seeded. The magnetic configuration is a reversed-field current sheet defined by
the flux function
\begin{equation}
  \begin{aligned}
    \psi_0(x,y) &= \frac{1}{2\pi}\,\tanh\!\Big(\frac{y}{h}\Big)\cos(\pi x)\sin(2\pi y),\\
    &\qquad h:\ 0.1 \to 1/300 .
  \end{aligned}
  \label{eq:ic}
\end{equation}
The hyperbolic tangent reverses the in-plane field $B_x=-\pp_y\psi_0$ across the
neutral line $y=0$ over a half-thickness $h$. The factor
$\cos(\pi x)\sin(2\pi y)$ sets the large-scale modulation and localizes the
layer near $x=0$. Close to the sheet,
$\psi_0\!\sim\!(y^2/h)\cos(\pi x)$, so $B_x$ crosses zero linearly at $y=0$ and
$J_z=\Lap\psi_0$ is concentrated there. The uniform guide field $g_0=1$ is
comparable to the peak reconnecting field, placing the system in a strong-guide,
order-unity plasma-$\beta$ regime. We sharpen the sheet from a broad,
well-resolved profile ($h=0.1$) to $h=1/300\approx3.3\times10^{-3}$, comparable
to the Sweet-Parker width $\delta_{\rm SP}\sim S^{-1/2}$
($h\approx1.5\,\delta_{\rm SP}$). Because the plasma starts from rest without
seeded islands, any later structure is generated by
Eqs.~\eqref{eq:continuity}-\eqref{eq:induction-g}.

We generate the reference solution of
Eqs.~\eqref{eq:continuity}-\eqref{eq:induction-g} with a conservative
finite-difference DNS on a uniform $N\times N$ nodal grid with $N=2049$. The
spacing is $\Delta x=1/(N-1)\approx4.9\times10^{-4}$, and the neutral line
$y=0$ lies exactly on a grid line. Spatial derivatives use second-order central
differences. We impose Eq.~\eqref{eq:bc} through parity reflection in one layer
of ghost nodes: even reflection enforces the Neumann conditions on $\rho$, $p$,
$g$, and the free-slip momentum components, while odd reflection enforces the
Dirichlet conditions on $u$, $v$, and $\psi$, together with
$\pp_{nn}\psi=0$, at the corresponding walls. After every stage, we also set
the Dirichlet nodes to zero: $u$ on the $x$ walls, $v$ on the $y$ walls, and
$\psi$ on all of $\partial\Omega$. Reconstructing $B_x$ and $B_y$ from $\psi$
through Eq.~\eqref{eq:Bfromflux} keeps the discrete magnetic field
divergence-free.

Time integration uses the classical fourth-order Runge-Kutta method. At each
step, $\Delta t$ is set by the stricter of the advective and diffusive
Courant-Friedrichs-Lewy (CFL) limits. The advective limit uses the fast
magnetosonic speed $c_f=\sqrt{c_s^2+|\bB|^2/\rho}$, with
$c_s^2=\gamma p/\rho$ and an upper limit of $0.30$. The diffusive limit is set
by $\max(\nu,\eta)$ with an upper limit of $0.20$. We add weak biharmonic
hyperviscosity, $-\nu_4\nabla_\perp^{4}$, to every evolved field to suppress
Nyquist-scale noise without damping resolved scales. Here
$\nu_4=\alpha\Delta x^{4}/\Delta t$ with $\alpha=0.01$, so the filter remains
confined to the grid scale. All fields are advanced in double precision to
$t=T=1.5$.
With these settings, the Sweet-Parker width spans
$\delta_{\rm SP}/\Delta x\approx4.6$ cells, and
the initial sheet spans $h/\Delta x\approx6.8$ cells. The reference grid
therefore resolves the reconnection layer marginally but consistently.

\begin{figure}[t]
\includegraphics[width=\linewidth]{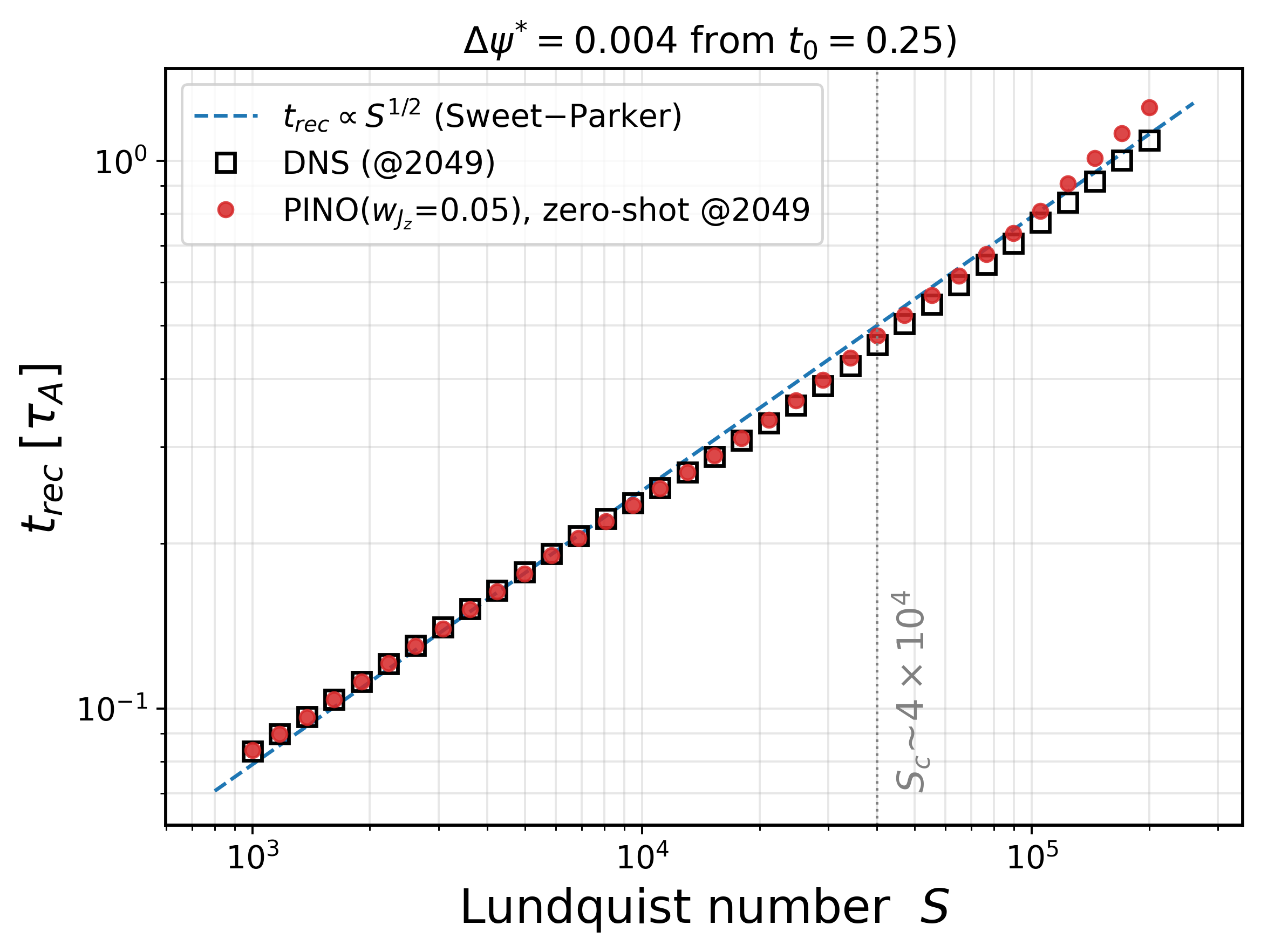}
\caption{\label{fig:huang-analog}Reference DNS configuration in the format of
Ref.~\cite{huang2010}: the reversed-field current sheet of
Eq.~\eqref{eq:ic} evolved under
Eqs.~\eqref{eq:continuity}-\eqref{eq:induction-g}, showing the formation of
the Sweet-Parker layer about the neutral line $y=0$.}
\end{figure}

\section{Physics-informed neural-operator surrogate}\label{sec:method}
We now describe our physics-informed neural operator (PINO), training
objectives, and diagnostics used for evaluation. Our PINO learns the
parametric solution map from the initial state and control parameters to the
fields defined in Sec.~\ref{sec:governing} at any query time.
\subsection{Architecture and structural constraints}\label{sec:arch}

Collecting the seven primitive fields in
$q=(\rho,u,v,w,p,\psi,g)$, the operator approximates
\begin{equation}
  \mathcal{G}_\theta:\;\big(q_0,\ \log_{10}S,\ t\big)\ \longmapsto\ \hat q(\cdot,t),
  \qquad t\in[0,T],
  \label{eq:operator}
\end{equation}
where $q_0=q(\cdot,0)$ is the initial condition in
Eqs.~\eqref{eq:ic-uniform}-\eqref{eq:ic}, and $S$ is the Lundquist number in
Sec.~\ref{sec:governing}. We query the operator directly at the requested time
instead of advancing it autoregressively, which avoids recursive error
accumulation \cite{mccabe2023stabilityautoregressiveneuraloperators}.
\paragraph{Input encoding.}

The input is an $11$-channel field on a uniform $n\times n$ grid with $n=513$,
obtained by strided coarsening of the DNS mesh in Sec.~\ref{sec:data}. Each
channel is a scalar field sampled on this grid. The seven initial-condition
channels are standardized separately as $\tilde q_0=(q_0-\mu)/\sigma$, using
the mean and standard deviation of each channel in the training set. The
Lundquist number is supplied as a constant channel, $\log_{10}S-4$, and the
query time as another constant channel, $2t/T-1\in[-1,1]$. Two coordinate
channels, $(x,y)$, provide absolute position and encode the wall geometry. The
operator returns the seven primitive fields at time $t$, which are mapped back
to physical units by inverse standardization.

\paragraph{Fourier neural operator.}
Our PINO is built around a two-dimensional Fourier neural operator
(FNO)~\cite{li_2021}. A standard convolution couples nearby grid points, whereas
an FNO performs its learned operations in Fourier space and couples the full
domain in each layer. A pointwise lifting map first converts the 11 input
channels into $w$ latent fields, collected in $h$. These fields then pass
through $L$ Fourier layers.

The core operation of each layer is a spectral convolution,
\begin{equation}
  (\mathcal{K}_\ell h)=\mathcal{F}^{-1}\!\big(R_\ell\odot\mathcal{F}h\big),
  \label{eq:spectralconv}
\end{equation}
where $\mathcal{F}$ and $\mathcal{F}^{-1}$ are the forward and inverse
two-dimensional discrete Fourier transforms. The layer transforms $h$ to
Fourier space, multiplies each retained wavenumber by learned weights, and then
transforms back; $\odot$ denotes multiplication over wavenumbers. Only the
lowest $K$ modes along each axis are retained. For every pair $(k_x,k_y)$,
$R_\ell\in\mathbb{C}^{\,w\times w\times K\times K}$ contains a $w\times w$
complex matrix that mixes the latent channels. All higher modes are set to zero.
The cutoff
$K=\min\!\big(K_{\max},\lfloor n/2\rfloor\big)$ never exceeds the Nyquist
wavenumber $\lfloor n/2\rfloor$ on an $n\times n$ grid. This restriction biases
the learned coupling toward the smooth, large-scale structure of the solution.

The spectral convolution in Eq.~\eqref{eq:spectralconv} cannot by itself
represent sharp, localized structures carried by discarded high wavenumbers.
Each Fourier layer therefore combines $\mathcal{K}_\ell$ with pointwise
operations. Residual connections add each branch back to its input, so the
layer learns a correction rather than a complete replacement:
\begin{eqnarray}
    h &\leftarrow \sigma\!\big(\mathcal{K}_\ell h + W_\ell\,h\big),\label{eq:fnolayer_eq1}\\[2pt]
h &\leftarrow \sigma\!\big(V^{(2)}_\ell\,\sigma\!\big(V^{(1)}_\ell h\big) + W'_\ell\,h\big).
  \label{eq:fnolayer_eq2}
\end{eqnarray}
Equation~\eqref{eq:fnolayer_eq1} adds the global spectral term
$\mathcal{K}_\ell h$ to the pointwise term $W_\ell h$. Here $W_\ell$ is a
$1\times1$ convolution: it mixes the $w$ channels independently at each grid
point but does not couple neighboring points. Equation~\eqref{eq:fnolayer_eq2}
adds a two-stage pointwise map. The map $V^{(1)}_\ell$ contracts the latent width
by a factor of two, and $V^{(2)}_\ell$ restores it; $W'_\ell h$ provides the
local residual. We use the Gaussian error linear unit (GELU) as the activation
$\sigma$~\cite{hendrycks2016gelu}. After $L$ layers, a final pointwise map projects the
latent fields onto the seven outputs. The models reported here use width
$w=64$, depth $L=4$, and $K_{\max}=48$ retained modes.

\paragraph{Domain padding for a non-periodic domain.}
The discrete Fourier transform treats the computational grid as periodic. Our
domain $\Omega$ is instead bounded by perfectly conducting walls, and values on
opposite walls generally differ. Applying the periodic transform directly would
join these unequal values, creating a discontinuity that appears in the
interior as spurious oscillations.

We reduce this artifact with reflection padding. Before applying the Fourier
layers, we mirror the input across each boundary by
$\lfloor\alpha n\rfloor$ cells per side, with $\alpha=1/16$. The Fourier stack
operates on the enlarged grid, and its output is cropped back to $\Omega$. The
reflection makes the field continuous across the original boundary, while the
remaining periodic join lies in a padded region that is discarded.

\paragraph{Structural constraint.}
The formulation in Sec.~\ref{sec:governing} provides one exact structural
constraint. Because $\mathcal{G}_\theta$ predicts $\psi$ rather than
$(B_x,B_y)$, Eq.~\eqref{eq:Bfromflux} reconstructs a solenoidal in-plane field,
and $\grad\!\cdot\!\bB=0$ holds for every prediction. The initial condition,
wall conditions in Eq.~\eqref{eq:bc}, and positivity of $\rho$ and $p$ are
enforced softly through the training objective in Sec.~\ref{sec:loss}. In
addition to supervision from DNS snapshots, including $t=0$, the objective
requires $\mathcal{G}_\theta(q_0,\log_{10}S,0)=q_0$ for every sampled initial
condition. This self consistency term anchors the operator to its exact input at
$t=0$.

\subsubsection{Training objective}\label{sec:loss}

We train the operator by supervised regression on the DNS ensemble and add a
strong-form MHD residual as a physics regularizer. The objective contains five
terms: a relative $L^2$ data loss $\Lloss_{L^2}$, a Sobolev ($H^1$) derivative
loss $\Lloss_{H^1}$, a current-density loss $\Lloss_{\rm Jz}$, an
initial-condition self-consistency loss $\Lloss_{\rm ic}$, and a strong-form
physics residual $\Lloss_{\rm phys}$:
\begin{equation}
  \Lloss \;=\; \Lloss_{L^2}\;+\;\lambda_{H^1}\,\Lloss_{H^1}
  \;+\;w_{\rm Jz}\,\Lloss_{\rm Jz}
  \;+\;w_{\rm ic}\,\Lloss_{\rm ic}
  \;+\;w_{\rm phys}\,\Lloss_{\rm phys},
  \label{eq:loss}
\end{equation}
where the coefficients $\lambda_{H^1}$, $w_{\rm Jz}$, $w_{\rm ic}$, and
$w_{\rm phys}$ set the relative weights of these terms.
\paragraph{Relative $L^2$ data term.}
The primary data term is the pointwise relative $L^2$ error, aggregated over the
seven standardized output channels and evaluated on the grid interior
$\Omega_{\rm int}$, excluding the boundary:
\begin{equation}
  \Lloss_{L^2}
  =\frac{\avg{\norm{\hat q-q}^2}{\Omega_{\rm int}}}
        {\avg{\norm{q}^2}{\Omega_{\rm int}}+\epsilon}.
  \label{eq:l2rel}
\end{equation}

\paragraph{Sobolev ($H^1$) derivative term.}
The pointwise $L^2$ term underweights low-amplitude velocity structures whose
signal is more apparent in their gradients than in their magnitudes. We
therefore add a separately normalized first-derivative term:
\begin{equation}
  \Lloss_{H^1}
  =\frac{\avg{\norm{\grad(\hat q-q)}^2}{\Omega_{\rm crop}}}
        {\avg{\norm{\grad q}^2}{\Omega_{\rm crop}}+\epsilon}.
  \label{eq:h1rel}
\end{equation}
Derivative-based terms such as $H^1$ loss, current-density loss, and strong-form residual use parity-aware spectral differentiation as described in
Sec.~\ref{sec:specdiff}. For each axis, $\rho$, $p$, $g$, $w$, and the
wall tangential velocity component are reflected evenly, while $\psi$ and the
wall-normal velocity component are reflected oddly. This produces a smooth,
periodic extension that can be differentiated in Fourier space without Gibbs
oscillations. It also matches the ghost-cell parity used by the DNS and reduces
the interior error in $\Lap\psi$ by more than an order of magnitude relative to
second-order differences. Strided coarsening leaves a small kink in the
even-parity extensions at the walls, and the predicted $\psi$ satisfies its odd
wall condition only to $\mathcal{O}(10^{-3})$. We therefore evaluate all
derivative-based losses on $\Omega_{\rm crop}$, obtained from
$\Omega_{\rm int}$ by removing a wall strip about eight cells wide. The
reconnection layer and outflows lie well inside this region. Normalizing
$\Lloss_{H^1}$ by the target gradient energy prevents derivatives of order
$\mathcal{O}(\Delta x^{-2})$ from overwhelming the pointwise loss and degrading
the smoother fields $\psi$ and $p$. We use $\lambda_{H^1}=0.5$.

\paragraph{Current density supervision.}
The out-of-plane current density $J_z=\Lap\psi$ is concentrated in the
reconnection layer and is difficult to predict because it is a second
derivative of $\psi$. We supervise it directly with a separately normalized
relative-$L^2$ loss on the Laplacian:
\begin{equation}
  \Lloss_{\rm Jz}
  =\frac{\avg{\norm{\Lap(\hat\psi-\psi)}^2}{\Omega_{\rm crop}}}
        {\avg{\norm{\Lap\psi}^2}{\Omega_{\rm crop}}+\epsilon}.
  \label{eq:jzrel}
\end{equation}
We use the same parity-aware spectral Laplacian and wall-strip crop as in the
$H^1$ term. The reported model uses $w_{\rm Jz}=5\times10^{-2}$; the ablation in
Sec.~\ref{sec:opt} also tests $w_{\rm Jz}=0$, $10^{-2}$, and $0.5$.

\paragraph{Initial condition.}
Because $q_0$ is only an input to $\mathcal{G}_\theta$, the architecture does
not force the operator to reproduce it at $t=0$. Labeled $t=0$ snapshots impose
this identity only for the initial conditions present in a batch. We therefore
query the operator again at $t=0$ for every sampled $(q_0,S)$ and penalize its
deviation from the supplied input:
\begin{equation}
  \Lloss_{\rm ic}
  =\Lloss_{L^2}\!\big(\hat q_0,q_0\big)+\lambda_{H^1}\,
  \Lloss_{H^1}\!\big(\hat q_0,q_0\big),\\
  \qquad \hat q_0\equiv\mathcal{G}_\theta\big(q_0,\log_{10}S,0\big).
  \label{eq:icrel}
\end{equation}
This is the sum of the relative-$L^2$ and $H^1$ functionals in
Eqs.~\eqref{eq:l2rel} and \eqref{eq:h1rel}, evaluated between the operator's
$t=0$ output and its input. It requires one additional forward pass per step and
explicitly enforces $\mathcal{G}_\theta(q_0,\log_{10}S,0)=q_0$. Because every
trajectory uses the same reference initial condition, this term is anchored to
one point in state space. It can be extended to a neighborhood of the
initial-condition manifold by applying smooth random perturbations to $q_0$ and
using the perturbed inputs as targets, but that option is disabled here. We set
$w_{\rm ic}=0.5$.

\paragraph{Strong-form physics residual.}
The physics loss evaluates the strong form residuals of
Eqs.~\eqref{eq:continuity}-\eqref{eq:induction-g} on the predicted physical
fields. Let $\R_i$ denote the residual of equation
$i\in\{\rho,m_x,m_y,m_z,p,\psi,g\}$, with additive terms $\R_i^{(k)}$ defined in
Appendix~\ref{app:residuals}. We normalize the mean-squared residual of each
equation by the mean-squared magnitude of its constituent terms:
\begin{equation}
  \Lloss_{\rm phys}
  =\frac{1}{7}\sum_i
   \frac{\avg{\R_i^2}{\Omega_{\rm crop}}}
        {\sum_k\avg{\big(\R_i^{(k)}\big)^2}{\Omega_{\rm crop}}+\epsilon}.
  \label{eq:phys}
\end{equation}
This makes each contribution dimensionless and of order unity when its equation
is not balanced. It also prevents the induction residual, which is about four
orders of magnitude smaller than the raw residuals of the wave-carrying
equations, from being hidden in the sum. Spatial derivatives use the same
parity-aware spectral method and wall crop as the $H^1$ and current-density
terms. The time derivative $\pp_t\hat q$ is obtained through forward-mode
automatic differentiation with respect to the continuous time input in
Eq.~\eqref{eq:operator}. We use the small weight $w_{\rm phys}=10^{-2}$ because
the derivative-aware $H^1$ and current density terms provide most of the
small accuracy at the operator resolution.

\subsubsection{Reconnection rate at $y=0$}\label{sec:reconny0}
We measure reconnection on the segment $|x|\le x_h$ of the neutral line $y=0$,
with $x_h=0.25$. The reconnected flux is the peak-to-peak variation of $\psi$
along this segment:
\begin{equation}
  \begin{aligned}
    \Psi_{\rm rec}(t)=
    &\max_{|x|\le x_h}\psi(x,0,t)
    -\min_{|x|\le x_h}\psi(x,0,t).
  \end{aligned}
  \label{eq:psirec}
\end{equation}
The instantaneous reconnection rate is the layer-averaged time derivative of
the flux function, equal to the mean out-of-plane reconnection electric field:
\begin{equation}
  \begin{aligned}
    \mathcal{R}_{\rm rec}(t)
    &=\big\langle\,\pp_t\psi\,\big\rangle_{|x|\le x_h,\,y=0}\\
    &=\big\langle\,{-}(u\,\pp_x\psi+v\,\pp_y\psi)+\eta\,\Lap\psi\,\big\rangle_{|x|\le x_h,\,y=0}.
  \end{aligned}
  \label{eq:rate}
\end{equation}
We also monitor the relative drift of the mean density $\langle\rho\rangle$ and
mean guide field $\langle g\rangle$. The impenetrable, perfectly conducting
walls conserve both quantities up to discretization error. These conservation
metrics are used only for evaluation, not as supervised training targets.

\subsubsection{Ablation}\label{sec:opt}

To isolate the effect of current density supervision, we sweep its weight over
$w_{\rm Jz}\in\{0,\,10^{-2},\,5\times10^{-2},\,0.5\}$ at fixed
architecture, grid, optimizer, schedule, and seed, with every other term of
Eq.~\eqref{eq:loss} held at the values of Sec.~\ref{sec:loss}. The arm
$w_{\rm Jz}=0$ omits current supervision, and $w_{\rm Jz}=5\times10^{-2}$ is
used for the results in Sec.~\ref{sec:results}. All four models are evaluated on
the same withheld Lundquist numbers and snapshot set, so the entries of
Table~\ref{tab:ablation-wjz} differ only through the weight.

We evaluate current density error in two ways. The first uses the parity-aware
spectral Laplacian from
Sec.~\ref{sec:specdiff}, denoted $J_z^{\rm spec}$, which is the operator
that enters both the training term of Eq.~\eqref{eq:jzrel} and the
reconnection diagnostic in Eq.~\eqref{eq:rate}. The second uses the
second-order central differences of Appendix~\ref{app:residuals}. The two
measures differ by about an order of magnitude in absolute value but show the
same trend.

The spectral current error falls by a factor of $2.7$ when $w_{\rm Jz}$
increases from $0$ to $10^{-2}$, dropping from $24.9\%$ to $9.3\%$, and by a
further factor of $1.4$ across the remaining two decades of weight, reaching
$6.4\%$ at $w_{\rm Jz}=0.5$. Most of the improvement therefore occurs as soon
as the term is introduced. Increasing the weight from $10^{-2}$ to $0.5$
recovers the final third of the improvement but steadily worsens other parts of
the objective.
The finite difference current error follows the same ordering, from $2.62$
to $1.67$ in relative $L^2$, but never approaches the spectral level, which
shows that the domain integrated finite difference metric is dominated by
grid scale content outside the layer rather than by the sheet itself.

The loss in accuracy occurs almost entirely in the flux function. Its relative $L^2$
error rises monotonically from $0.82\%$ at $w_{\rm Jz}=0$ to $1.18\%$ at
$w_{\rm Jz}=0.5$, a degradation of $44\%$ in relative terms, consistent with
the redistribution of error toward the separatrices and current sheet. The
remaining primitive fields are largely insensitive to the weight. Density,
pressure, and guide-field errors change by less than $3\%$ of
their own error across the full sweep, and the three velocity components are
within $2\%$ of one another. The choice of $w_{\rm Jz}$ therefore does not
trade current accuracy against the bulk compressible response. The total
validation loss increases by $7.6\%$ across the sweep, from
$7.48\times10^{-4}$ to $8.05\times10^{-4}$, mainly because of the increase in
$\psi$ error.

The strong form residuals in Eq.~\eqref{eq:phys-norm}, evaluated on the same
validation set, show where this tradeoff appears. The continuity, pressure,
and guide field residuals remain fixed at $0.37$, $0.51$, and $0.38$ in every
case, consistent with a floor set by the snapshot cadence rather than the
objective.
The balances that are free of the floor all degrade with $w_{\rm Jz}$: the
in-plane momentum residuals grow from $0.020$ to $0.071$ and from $0.171$ to
$0.288$, the out-of-plane momentum residual from $1.8\times10^{-3}$ to
$6.1\times10^{-3}$, and the induction residual, the one governing
reconnection, from $0.028$ to $0.042$. Current supervision and the strong-form
residual are therefore not aligned. Forcing $\Lap\hat\psi$ toward the DNS
current sharpens the layer but weakens pointwise satisfaction of the governing
equations; the $50\%$ increase in $\R_\psi$ measures this tension directly. The
initial-condition self-consistency errors in Eq.~\eqref{eq:icrel} change by a
similar fraction. The largest shift is in the in-plane normal velocity, from
$0.146$ to $0.155$, which indicates that the $t=0$ anchor is not the source of
the tradeoff.

\begin{figure*}[t]
\includegraphics[width=\linewidth]{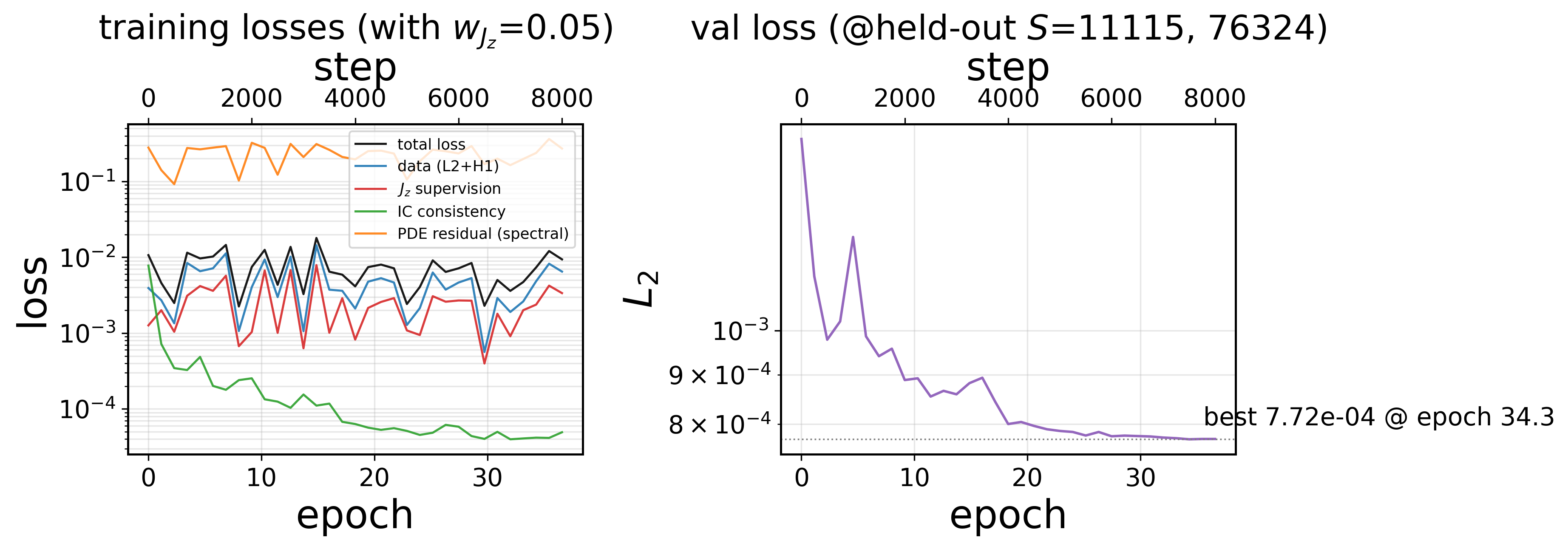}
\caption{\label{fig:loss}Training and validation loss histories for the
objective of Eq.~\eqref{eq:loss} at $w_{\rm Jz}=5\times10^{-2}$, the
intermediate arm of Table~\ref{tab:ablation-wjz}.}
\end{figure*}

\begin{table}[t]
\caption{\label{tab:ablation-wjz}Ablation of the current-density
supervision weight $w_{\rm Jz}$ of Eq.~\eqref{eq:jzrel}, at fixed
architecture, grid, optimizer, and seed. Upper block: relative $L^2$ errors
on the withheld Lundquist numbers, with $J_z^{\rm spec}$ evaluated by the
parity-aware spectral Laplacian of Sec.~\ref{sec:specdiff} and $J_z$ by
second-order central differences. Lower block: per equation normalized
strong-form residuals of Eq.~\eqref{eq:phys-norm} on the same validation
set.}
\begin{ruledtabular}
\begin{tabular}{lcccc}
 & \multicolumn{4}{c}{$w_{\rm Jz}$}\\
\cline{2-5}
Quantity & $0$ & $10^{-2}$ & $5\times10^{-2}$ & $0.5$ \\
\hline
\multicolumn{5}{l}{\textit{Relative $L^{2}$ error}}\\
$J_z^{\rm spec}$ & 0.249 & 0.093 & 0.075 & 0.064 \\
$J_z$            & 2.618 & 2.269 & 2.096 & 1.671 \\
$\psi$           & 0.0082 & 0.0090 & 0.0102 & 0.0118 \\
$\rho$           & 0.0012 & 0.0012 & 0.0012 & 0.0012 \\
$p$              & 0.0020 & 0.0020 & 0.0020 & 0.0020 \\
$g$              & 0.0015 & 0.0015 & 0.0015 & 0.0015 \\
$u$              & 0.0322 & 0.0322 & 0.0323 & 0.0327 \\
$v$              & 0.0378 & 0.0379 & 0.0379 & 0.0383 \\
$w$              & 0.0390 & 0.0390 & 0.0390 & 0.0393 \\
Validation loss  & $7.48\times10^{-4}$ & $7.57\times10^{-4}$
                 & $7.72\times10^{-4}$ & $8.05\times10^{-4}$ \\
\hline
\multicolumn{5}{l}{\textit{Normalized strong-form residual}}\\
$\R_\rho$  & 0.370 & 0.369 & 0.369 & 0.369 \\
$\R_{m_x}$ & 0.020 & 0.044 & 0.053 & 0.071 \\
$\R_{m_y}$ & 0.171 & 0.232 & 0.243 & 0.288 \\
$\R_{m_z}$ & 0.0018 & 0.0038 & 0.0045 & 0.0061 \\
$\R_p$     & 0.512 & 0.512 & 0.512 & 0.512 \\
$\R_\psi$  & 0.028 & 0.031 & 0.035 & 0.042 \\
$\R_g$     & 0.378 & 0.379 & 0.379 & 0.379 \\
\end{tabular}
\end{ruledtabular}
\end{table}

\section{Results and discussion}\label{sec:results}

We evaluate the surrogate against the DNS described in Sec.~\ref{sec:data}. All
reported results use $w_{\rm Jz}=5\times10^{-2}$, the operating point selected
in Sec.~\ref{sec:opt}. Unless a single time is stated, errors are reported as
$\lVert\hat q-q\rVert_2/\lVert q\rVert_2$ and averaged over $t>0$. We withhold
two Lundquist numbers, $S=1.1\times10^{4}$ and $S=7.6\times10^{4}$, to test
interpolation; the other 32 values span
$10^{3}\le S\le2\times10^{5}$. Field comparisons use the $513^2$ training grid.
As shown in Sec.~\ref{sec:res-superres}, querying the same operator on grids
from $257^2$ to the $2049^2$ DNS mesh changes the primitive-field errors by only
a few tenths of a percentage point. Reconnection rates are evaluated on the
$2049^2$ mesh, as explained in Sec.~\ref{sec:res-rate}.

\begin{figure*}[t]
\includegraphics[width=\linewidth]{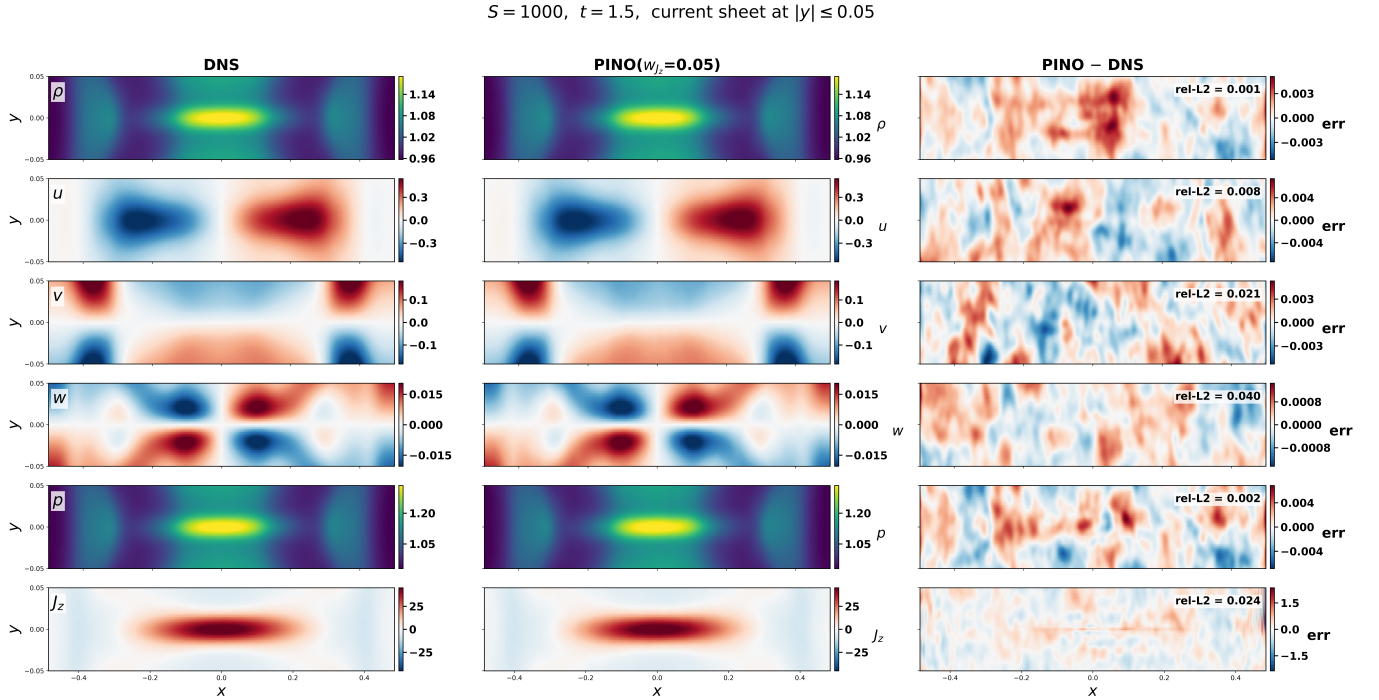}
\caption{\label{fig:strip-S1e3-t15}DNS (left), PINO prediction (center), and
difference (right) within the current sheet strip $|y|\le0.05$ at
$S=10^{3}$, $t=1.5$, for the fields $\rho$, $u$, $v$, $w$, $p$, and
$J_z=\Lap\psi$ evaluated with the spectral Laplacian of
Sec.~\ref{sec:specdiff}. Per panel relative $L^2$ errors are annotated.}
\end{figure*}

\begin{figure*}[t]
\includegraphics[width=\linewidth]{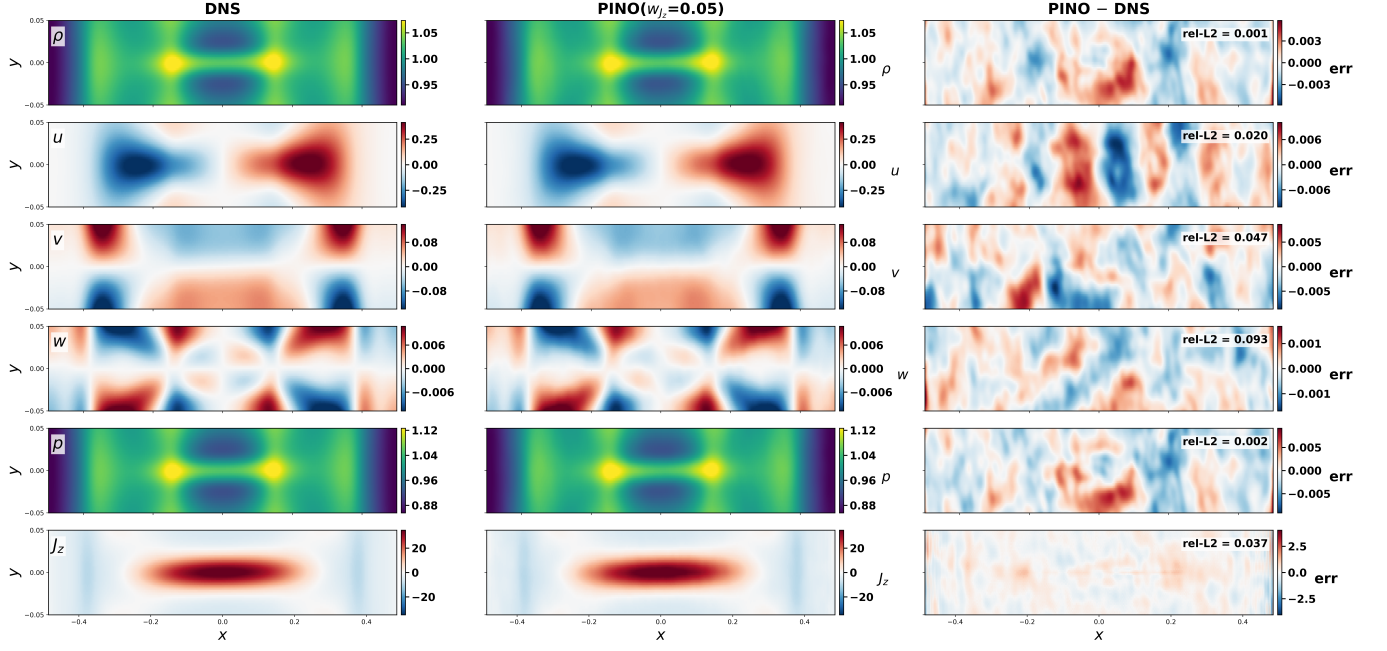}
\caption{\label{fig:strip-S1e3-t19}Same as
Fig.~\ref{fig:strip-S1e3-t15}, but at $t=1.9$ during the quasi-steady phase.}
\end{figure*}

\begin{figure*}[t]
\includegraphics[width=\linewidth]{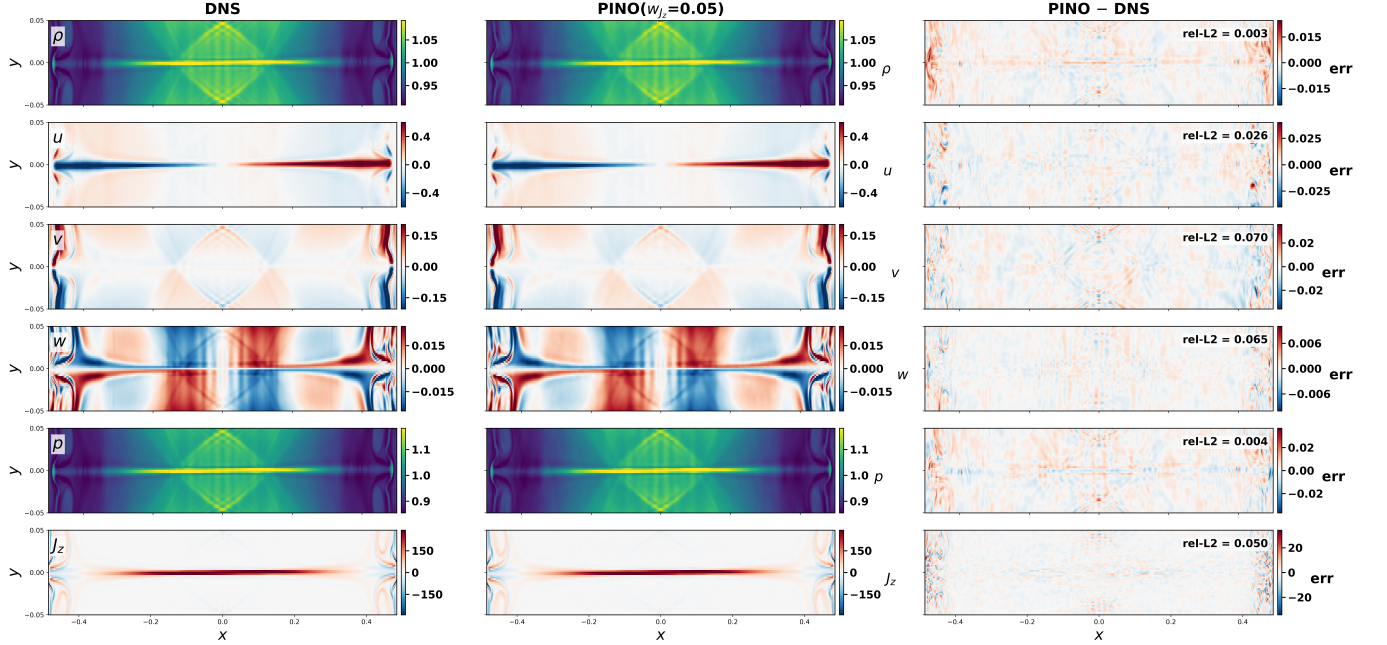}
\caption{\label{fig:strip-S2e5-t15}Same as
Fig.~\ref{fig:strip-S1e3-t15}, but at $S=2\times10^{5}$ and $t=1.5$, where
the layer is thinnest and the current-density error is largest.}
\end{figure*}

\begin{figure*}[t]
\includegraphics[width=\linewidth]{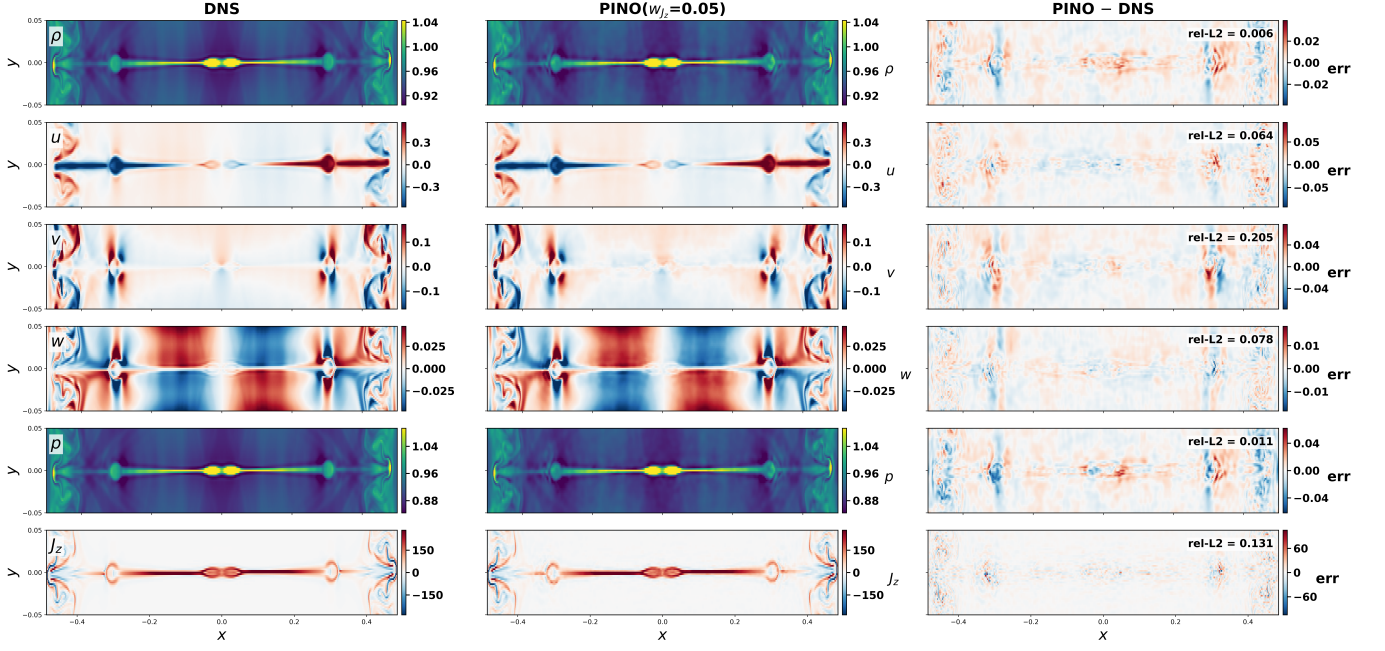}
\caption{\label{fig:strip-S2e5-t19}Same as
Fig.~\ref{fig:strip-S2e5-t15}, but at $t=1.9$ during the plasmoid stage.}
\end{figure*}

\subsection{Fields in the reconnection layer}\label{sec:res-fields}

Figures~\ref{fig:strip-S1e3-t15}-\ref{fig:strip-S2e5-t19} compare the
surrogate and DNS within the current sheet strip $|y|\le0.05$ at both ends of
the Lundquist-number sweep and at two times. Because every diagnostic in
Sec.~\ref{sec:reconny0} samples this region, we report strip errors rather than
full-domain averages dominated by the large ideal regions. At $S=10^{3}$ and
$t=1.5$, every field is accurate to within a few percent: $0.13\%$ for density,
$0.18\%$ for pressure, $0.8\%$ for the outflow velocity $u$, $2.1\%$ for the
inflow velocity $v$, and $2.4\%$ for the current density. By $t=1.9$, the errors
in $u$, $v$, and $J_z$ rise to $2.0\%$, $4.8\%$, and $3.7\%$, respectively. The
out-of-plane velocity has the largest relative error, $9.3\%$, because its
amplitude is only a few times $10^{-2}$.

At $S=2\times10^{5}$ and $t=1.5$, the strip errors increase to $2.7\%$ in $u$,
$7.0\%$ in $v$, and $5.0\%$ in $J_z$, while density and pressure remain below
$0.5\%$. Figure~\ref{fig:strip-S2e5-t15} shows that the surrogate resolves the
thin layer instead of smoothing it away. The predicted current ridge reaches
the DNS peak of about $200$, and the Alfv\'enic outflow jets and surrounding
shock lattice are present. Most of the remaining error lies near the outflow
tips and separatrices. At $t=1.9$, the DNS has entered the plasmoid stage and
contains a chain of secondary islands. The surrogate captures the number and
amplitudes of the islands, but the outer islands are slightly displaced. A
pointwise norm counts this phase shift both where an island is absent and where
it is misplaced, raising the strip errors in $v$ and $J_z$ to $20\%$ and
$13\%$; the error in $u$ is $6.4\%$.

Across the full domain and trajectory for the two withheld Lundquist numbers,
the surrogate predicts density, pressure, and guide field to
$0.12\%$, $0.20\%$, and $0.15\%$, the velocity components to $3.2$, $3.8$,
and $3.9\%$, the flux function to $1.0\%$, and the spectral current density
to $7.5\%$ (the $w_{\rm Jz}=5\times10^{-2}$ column of
Table~\ref{tab:ablation-wjz}). These errors are comparable to or smaller than
those reported for Fourier neural operators applied to periodic incompressible
MHD~\cite{rosofsky2023} and tokamak-edge dynamics~\cite{gopakumar2024}, despite
the wall-bounded, compressible setting and the two-decade sweep in $S$.

\begin{figure*}[t]
\includegraphics[width=\linewidth]{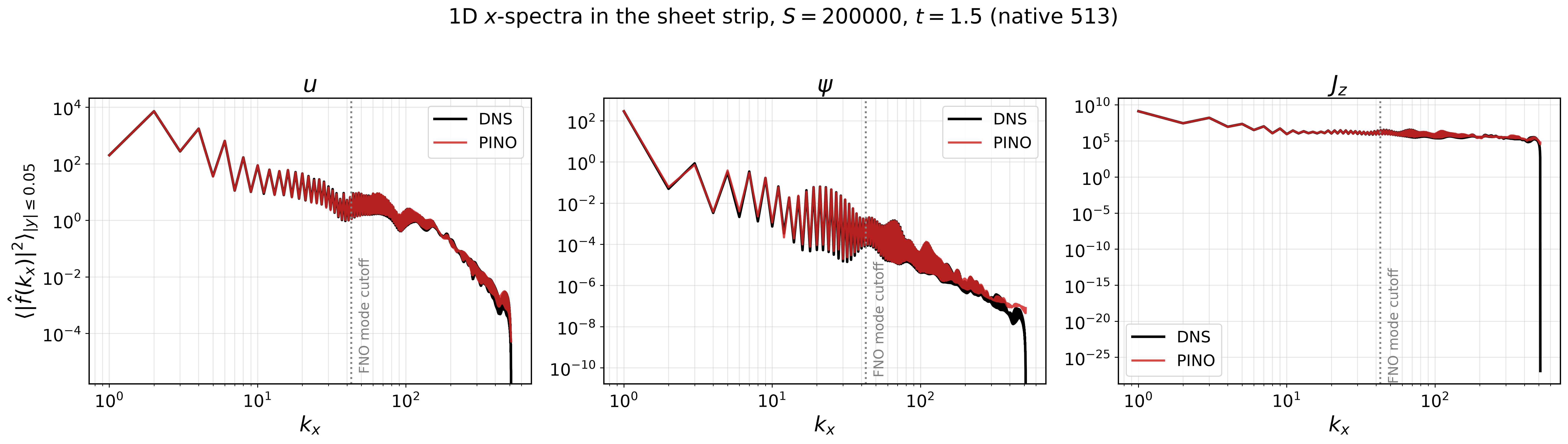}
\caption{\label{fig:spectra}One-dimensional spectra of $u$, $\psi$, and
$J_z$, averaged over the sheet strip $|y|\le0.05$, at $S=2\times10^{5}$,
$t=1.5$, on the native $513^2$ grid. The dotted line marks the operator
mode cutoff of Sec.~\ref{sec:arch} at unit domain index $\approx43$.}
\end{figure*}

Figure~\ref{fig:spectra} compares the same fields in wavenumber space. The
panels show one-dimensional $x$ spectra averaged over the strip at
$S=2\times10^{5}$, $t=1.5$, for $u$, $\psi$, and $J_z$, with the operator
cutoff of Sec.~\ref{sec:arch} marked at unit domain index
$K_{\max}/(1+2\alpha)\approx43$. The surrogate tracks the DNS spectrum
through the energy-carrying range and beyond the cutoff, including the nearly
flat $J_z$ spectrum of the thin sheet up to its grid-scale dissipation range.
Although the learned spectral weights stop at $k\approx43$, the pointwise
channel mixing and activation nonlinearity in
Eqs.~\eqref{eq:fnolayer_eq1} and \eqref{eq:fnolayer_eq2} generate high
wavenumber content. The current-density and $H^1$ losses calibrate this content
against the DNS. The main mismatch is a small excess of energy just above the
cutoff, corresponding to the fine-scale speckle in the error maps of
Figs.~\ref{fig:strip-S2e5-t15} and \ref{fig:strip-S2e5-t19}. The mode
cutoff marks where fidelity begins to degrade, not where the prediction ends.

\subsection{Reconnection rate across the Lundquist sweep}\label{sec:res-rate}

The layer-averaged rate $\mathcal{R}_{\rm rec}(t)$ in Eq.~\eqref{eq:rate}
provides a direct physics-level test of the surrogate. Near the X point, the
advective contribution vanishes with the flow, and the rate is set by
$\eta\Lap\psi$. It therefore depends on the current density precisely where the
fields are hardest to represent.

The numerical estimator must also be accurate. When Eq.~\eqref{eq:rate} is
evaluated by finite differences on the $513^2$ grid, even the DNS fields differ
from the solver's internally recorded rate by as much as $14\%$ at
$S=2\times10^{5}$. This is diagnostic discretization error, not model error. We
therefore query the surrogate zero-shot on the $2049^2$ solver mesh and compare
it directly with the solver's rate history. Section~\ref{sec:res-superres}
validates this resolution transfer.

\begin{figure*}[t]
\includegraphics[width=\linewidth]{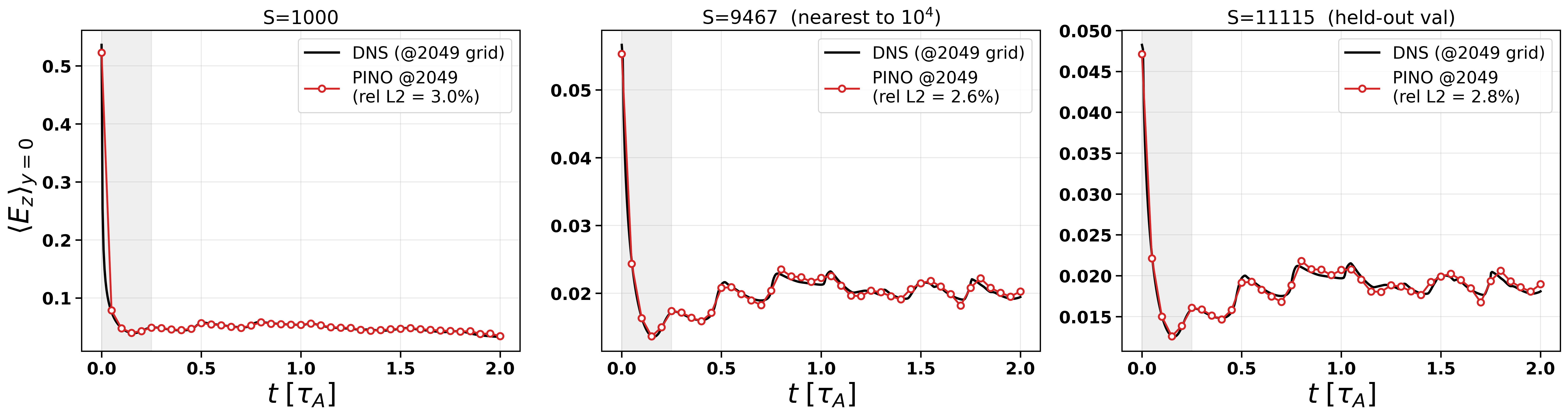}
\caption{\label{fig:rate-polished}Layer-averaged reconnection rate
$\mathcal{R}_{\rm rec}(t)$ of Eq.~\eqref{eq:rate} for the DNS (black,
solver rate history at $2049^2$) and the surrogate queried zero-shot on the
same mesh (red), at $S=10^{3}$, at $S=9467$ (the trained value nearest
$10^{4}$), and at the withheld $S=1.1\times10^{4}$. Legends quote the
relative $L^{2}$ error over $t>0.25$; the shaded band marks the excluded
initial transient.}
\end{figure*}

Figure~\ref{fig:rate-polished} compares the rate at $S=10^{3}$, at the trained
value $S=9467$ nearest $10^{4}$, and at the withheld value
$S=1.1\times10^{4}$. Over $t>0.25$, the relative errors are $3.0\%$, $2.6\%$,
and $2.8\%$, respectively. The surrogate captures the initial relaxation,
oscillatory settling, and quasi-steady plateau. Its accuracy at the withheld
value matches that at the neighboring training values.

Across the full sweep, the flow regime matters more than whether a value was
included in training. The error stays below $3\%$ up to
$S=1.1\times10^{4}$, grows gradually from $3.1$ to $7.8\%$ over
$1.3\times10^{4}\le S\le4.7\times10^{4}$, and then breaks upward: $9.2\%$
at $S=5.5\times10^{4}$, then $13$ to $18\%$ for
$6.5\times10^{4}\le S\le9.0\times10^{4}$. The withheld value
$S=7.6\times10^{4}$ has an error of $15.2\%$, consistent with its trained
neighbors. For $S\ge1.05\times10^{5}$, the error reaches $32$-$42\%$. This
sharp increase coincides with the plasmoid transition in the DNS. Above
$S\approx10^{5}$, the DNS develops a reconnection burst near $t=1.8$ that is
still growing when the run ends at $T=2$. The surrogate reproduces the preceding
plateau but underestimates the burst, whose developed stage is absent from the
training trajectories. The high-$S$ error therefore reflects the limited time
window as well as the model.

The time window also limits the scaling comparison with
Ref.~\cite{huang2010}. Figure~\ref{fig:huang-analog} plots the
reconnection time $t_{\rm rec}$, defined as the interval, measured from
$t_0=0.25$, over which the time integral of $\mathcal{R}_{\rm rec}$
accumulates a fixed flux increment $\Delta\psi^{*}=4\times10^{-3}$,
against $S$ on logarithmic axes. A power-law fit gives an exponent of $0.478$
for the DNS and $0.497$ for the surrogate, both consistent with the Sweet
Parker exponent of $1/2$. The surrogate matches the DNS reconnection time with
a median error of $1.0\%$ and a maximum error of $14.8\%$ at
$S=2\times10^{5}$, where it reconnects slightly too slowly. Neither series shows
the plateau reported above $S_c\sim4\times10^{4}$ in Ref.~\cite{huang2010}.
For $S\ge10^{5}$, the flux threshold is crossed between $t=1.0$ and $t=1.35$,
before the burst begins, so the diagnostic samples only the laminar
Sweet-Parker phase. Observing the plateau requires reference runs several
Alfv\'en times longer. The reconnected flux $\Psi_{\rm rec}$ in
Eq.~\eqref{eq:psirec} is subject to a stricter amplitude floor: at
$S=2\times10^{5}$ the flux reconnected inside the window is a few times
$10^{-3}$, comparable to the surrogate's absolute accuracy on $\psi$, so
the pointwise excursion cannot be resolved reliably there.

\subsection{Resolution transfer}\label{sec:res-superres}

The Fourier stack is nominally discretization-invariant because its learned
weights act on wavenumbers rather than grid points. In practice, every
grid dependent part of the architecture must scale with resolution. Here that
part is the reflection padding from Sec.~\ref{sec:arch}, specified as the fixed
fraction $\alpha=1/16$. Each learned spectral weight corresponds to a physical
wavelength on the padded domain. Holding the padding at a fixed number of pixels
while changing the grid would change the padded domain length and shift every
learned wavelength. In our tests, fixed pixel padding increased the velocity
error by an order of magnitude when the resolution doubled; fractional padding
did not. All results below use fractional padding.

\begin{figure*}[t]
\includegraphics[width=\linewidth]{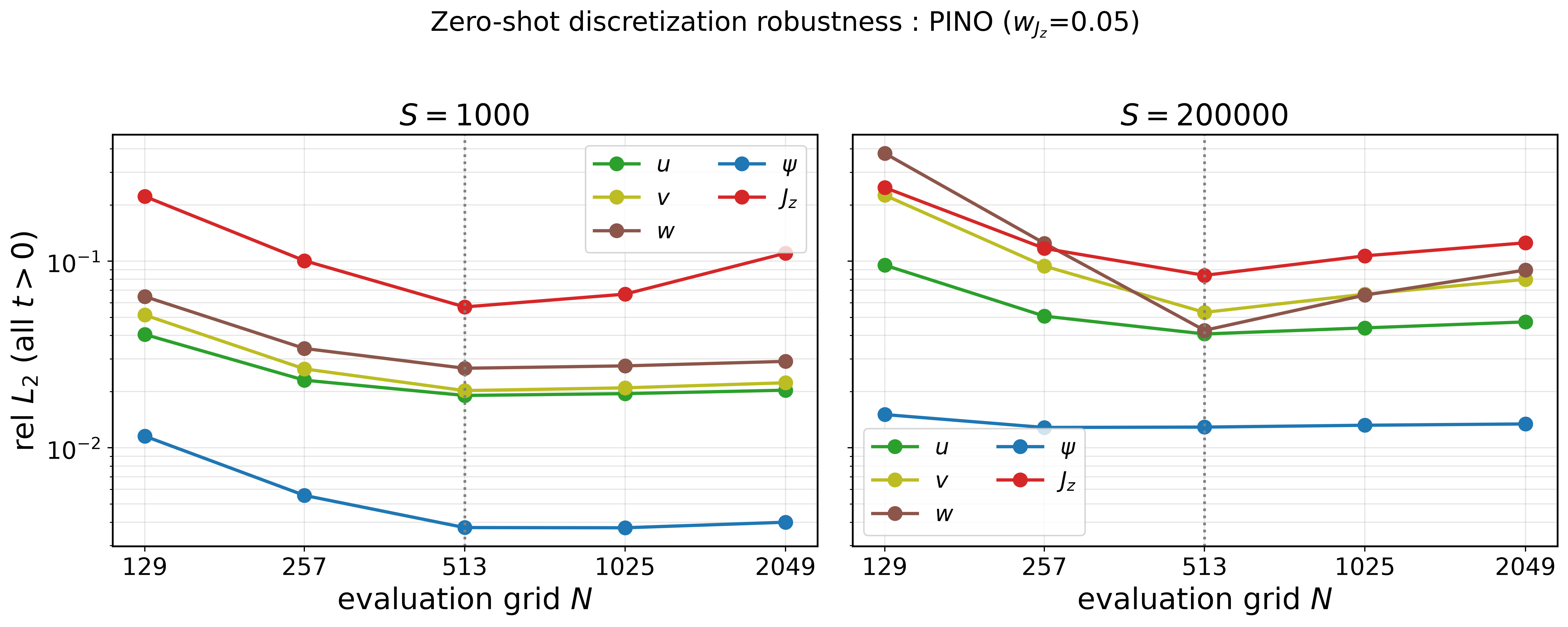}
\caption{\label{fig:gridsweep}Trajectory-averaged relative $L^{2}$ error
versus evaluation grid for the operator trained only at $513^2$ (dotted
line), queried zero-shot with the fractional pad $\alpha=1/16$, at
$S=10^{3}$ (left) and $S=2\times10^{5}$ (right). Trained at 513 grid size.}
\end{figure*}

Figure~\ref{fig:gridsweep} and Table~\ref{tab:transfer} quantify the transfer.
Although trained only at $513^2$, the operator maintains nearly constant
velocity error at $S=2\times10^{5}$: $4.1\%$ at $513^2$, $4.4\%$ at $1025^2$,
and $4.7\%$ at $2049^2$. The number of query points increases sixteenfold. The
flux function error changes only from $1.29\%$ to $1.34\%$. Current density is
more sensitive, increasing from $8.4\%$ to $12.5\%$. At $S=10^{3}$, the errors
are smaller: the error in $u$ changes from $1.9\%$ to $2.0\%$, while the error
in $\psi$ remains near $0.4\%$. Errors increase more rapidly on coarser grids,
reaching $9.5\%$ for $u$ and $25\%$ for current density at $129^2$ and
$S=2\times10^{5}$. At that resolution, however, the strided DNS reference also
aliases the sheet because its width lies below the coarse-grid Nyquist scale.

\begin{table}[t]
\caption{\label{tab:transfer}Relative $L^{2}$ error (\%) at
$S=2\times10^{5}$, averaged over the trajectory, for the operator trained
at $513^2$ and queried zero-shot on the stated grids with the fractional
pad $\alpha=1/16$. $J_z^{\rm spec}$ uses the parity-aware spectral
Laplacian of Sec.~\ref{sec:specdiff}, with the wall strip crop scaled in
proportion to the grid.}
\begin{ruledtabular}
\begin{tabular}{lccc}
Field & $513^2$ & $1025^2$ & $2049^2$ \\
\hline
$u$      & 4.1 & 4.4 & 4.7 \\
$v$      & 5.3 & 6.7 & 8.0 \\
$w$      & 4.3 & 6.6 & 9.0 \\
$\psi$   & 1.29 & 1.32 & 1.34 \\
$J_z^{\rm spec}$ & 8.4 & 10.7 & 12.5 \\
\end{tabular}
\end{ruledtabular}
\end{table}

These results justify the procedure used in Sec.~\ref{sec:res-rate}:
derivative-based diagnostics should be computed by querying the surrogate on the
DNS mesh, where finite difference bias is negligible, rather than on the
training grid. A $2049^2$ snapshot takes less than one second on one A100 GPU,
and a 41 snapshot trajectory takes about 30 seconds, compared with several hours
for the DNS. Even on the solver grid, the surrogate is more than two orders of
magnitude faster per trajectory.

\section{Conclusions}\label{sec:conclusions}
 
We developed a physics-informed neural operator for two-dimensional,
compressible, viscous, resistive MHD reconnection in a wall-bounded domain. The
operator is conditioned on the initial state, Lundquist number, and continuous
query time. Predicting the flux function makes $\grad\!\cdot\!\bB=0$ exact,
direct-time queries avoid autoregressive error accumulation, and parity-aware
spectral differentiation matches the wall treatment used by the DNS. We tested
the model over $10^{3}\le S\le2\times10^{5}$, withholding two values of $S$
from training.

For the withheld cases, the trajectory-averaged relative errors are $0.12\%$,
$0.20\%$, and $0.15\%$ in density, pressure, and guide field; $3.2$-$3.9\%$ in
the three velocity components; $1.0\%$ in the flux function; and $7.5\%$ in the
spectral current density. Within the current sheet strip at
$S=2\times10^{5}$ and $t=1.5$, the errors are $2.7\%$ in the outflow, $7.0\%$
in the inflow, and $5.0\%$ in the current. The model resolves the current ridge,
Alfv\'enic jets, shock lattice, and spectral content beyond the retained-mode
cutoff. At $t=1.9$, phase shifts in the outer secondary islands raise the errors
in $v$ and $J_z$ to $20\%$ and $13\%$.

Accurate $\psi$ alone does not guarantee accurate current density. Without
current supervision, the error in $\psi$ is $0.82\%$, but the error in
$\Lap\psi$ is $24.9\%$. Adding the current loss reduces this error to $9.3\%$ at
$w_{\rm Jz}=10^{-2}$ and $6.4\%$ at $w_{\rm Jz}=0.5$, while increasing the
flux function error to $1.18\%$ and the induction residual from $0.028$ to
$0.042$. We therefore use the intermediate weight
$w_{\rm Jz}=5\times10^{-2}$. The remaining sheet scale error appears to be
limited by the 48 retained modes rather than by the loss weight.

The reconnection rate error remains below $3\%$ through
$S=1.1\times10^{4}$ and reaches $7.8\%$ at $S=4.7\times10^{4}$. It rises to
$32$-$42\%$ for $S\ge1.05\times10^{5}$, where the DNS develops a late
plasmoid mediated burst that is not fully contained in the training window. The
reconnection time scaling remains close to Sweet-Parker: the fitted exponent is
$0.497$ for the surrogate and $0.478$ for the DNS. The surrogate matches the DNS
reconnection time with a median error of $1.0\%$ and a maximum error of $14.8\%$
at $S=2\times10^{5}$.

Resolution transfer is reliable when reflection padding scales as a fraction of
the grid. From $513^2$ to $2049^2$, the velocity error at
$S=2\times10^{5}$ changes from $4.1\%$ to $4.7\%$, the flux function error from
$1.29\%$ to $1.34\%$, and the current density error from $8.4\%$ to $12.5\%$.
A 41 snapshot trajectory on the $2049^2$ grid takes about 30 seconds on one A100
GPU, compared with several hours for the DNS.

% ======================================================================
\begin{acknowledgments}
A.~Kumar's simulations presented in this paper were performed on the MIT-PSFC
partition of the Engaging cluster at the MGHPCC facility
(\url{https://www.mghpcc.org}), which was funded by DOE grant number
DE-FG02-91-ER54109. A.~Kumar also thanks the MIT Office of Research Computing
and Data for access to the ORCD high-performance computing facility, on which
the reference simulations, selected surrogate training checks, and inference
calculations were performed.
\end{acknowledgments}

\section*{Data availability}
The source code for the reference solver, the neural operator surrogate,
and the training and evaluation pipelines, together with the trained model
checkpoints and the scripts that regenerate every figure in this paper,
are available from the corresponding author upon reasonable request and
will be deposited in a public repository upon publication.

% ======================================================================
\appendix
\section{Strong-form residuals and their normalization}\label{app:residuals}

The physics term of the training objective measures how well the predicted
fields satisfy the governing equations pointwise. This appendix defines those
residuals, the per-equation normalization applied to them, and the numerical
conventions used to evaluate them.

The surrogate outputs the primitive state $q=(\rho,u,v,w,p,\psi,g)$, consisting
of the mass density $\rho$, the velocity components $(u,v,w)$, the thermal
pressure $p$, the in-plane magnetic flux function $\psi$, and the out-of-plane
guide field $g$. From these we form the momentum components
$m_x=\rho u$, $m_y=\rho v$, and $m_z=\rho w$; the in-plane magnetic field
$B_x=-\pp_y\psi$ and $B_y=\pp_x\psi$; the out-of-plane current density
$J_z=\Lap\psi$ with $\Lap=\pp_x^2+\pp_y^2$; and the total pressure
$P_T=p+\tfrac12(B_x^2+B_y^2+g^2)$. The transport coefficients are the
kinematic viscosity $\nu$ and the resistivity $\eta=1/S$, with adiabatic index
$\gamma=5/3$ and magnetic Prandtl number $\nu/\eta=1$.

Writing each of the seven evolution equations with all terms moved to one side,
and recovering the primitive velocities from the predicted momenta through
$u=m_x/\rho$ and its counterparts, gives the residuals
\begin{align}
  \R_\rho  &= \pp_t\rho+\pp_x m_x+\pp_y m_y,\notag\\
  \R_{m_x} &= \pp_t m_x+\pp_x\!\big(m_x u+P_T-B_x^2\big)\notag\\
           &\qquad +\pp_y\!\big(m_x v-B_xB_y\big)-\nu\Lap m_x,\notag\\
  \R_{m_y} &= \pp_t m_y+\pp_x\!\big(m_y u-B_xB_y\big)\notag\\
           &\qquad +\pp_y\!\big(m_y v+P_T-B_y^2\big)-\nu\Lap m_y,\notag\\
  \R_{m_z} &= \pp_t m_z+\pp_x\!\big(m_z u-gB_x\big)
              +\pp_y\!\big(m_z v-gB_y\big)-\nu\Lap m_z,\notag\\
  \R_p     &= \pp_t p+u\,\pp_x p+v\,\pp_y p+\gamma\,p\,(\pp_x u+\pp_y v),\notag\\
  \R_\psi  &= \pp_t\psi+u\,\pp_x\psi+v\,\pp_y\psi-\eta\Lap\psi,\notag\\
  \R_g     &= \pp_t g+\pp_x\!\big(u g-wB_x\big)+\pp_y\!\big(v g-wB_y\big)-\eta\Lap g,
  \label{eq:residuals}
\end{align}
A residual that vanishes identically means the corresponding equation is
satisfied exactly at that point. Spatial derivatives are
evaluated by second-order central differences on the operator grid. The time
derivative $\pp_t\hat q$ is obtained through forward-mode automatic
differentiation with respect to the continuous time input of
$\mathcal{G}_\theta$, rather than by differencing snapshots. When forming the
residuals we exclude the two outermost cells on each side, so that no one-sided
boundary stencil enters; the spatial means defined below are taken over the
further-cropped region $\Omega_{\rm crop}$. The reference solver imposes reflection
symmetry at the walls and neutral line through parity matched stencils; the
surrogate learns these symmetries from the supervised data.

The seven residuals in Eq.~\eqref{eq:residuals} carry different physical
dimensions and differ in raw magnitude by several orders of magnitude: the
induction residual $\R_\psi$ is roughly $10^{4}$ times smaller than the
residuals of the wave-carrying equations. Summing them unweighted would let the
largest ones dominate and would hide exactly the balance that governs
reconnection. We therefore normalize each residual by the mean-squared
magnitude of its own additive terms. Let $\R_i^{(k)}$ denote the $k$th additive
term of $\R_i$, that is, one of the summands appearing on the right-hand side
of the corresponding line of Eq.~\eqref{eq:residuals}. For the $x$-momentum
equation these terms are
\begin{eqnarray}
  \R_{m_x}^{(k)}\in
  \big\{\pp_t m_x,\ \pp_x(m_x u+P_T-B_x^2), \nonumber \\ \
        \pp_y(m_x v-B_xB_y),\ \nu\Lap m_x\big\},
  \label{eq:terms-mx}
\end{eqnarray}
and the remaining equations are decomposed in the same way. The normalized
per-equation residual is then
\begin{equation}
  \widetilde{\R}_i
   =\frac{\big\langle\,\R_i^2\,\big\rangle_{\Omega_{\rm crop}}}
        {\sum_{k}\big\langle\,(\R_i^{(k)})^2\,\big\rangle_{\Omega_{\rm crop}}+\epsilon},
  \label{eq:phys-norm}
\end{equation}
where $\langle\cdot\rangle_{\Omega_{\rm crop}}$ is a spatial mean over the grid
interior with a wall strip about eight cells wide removed, and $\epsilon$ is a
small stabilizing constant. Each $\widetilde{\R}_i$ is dimensionless and of
order unity when its equation is not balanced, so the seven equations
contribute comparably. The physics term of the training objective is the mean
of these seven quantities, and it carries the weight $w_{\rm phys}=10^{-2}$.
The quantities reported in the lower block of Table~\ref{tab:ablation-wjz} are
the individual $\widetilde{\R}_i$, listed there simply as $\R_i$.

The current-density supervision that accompanies this term is normalized in the
same spirit, dividing the mean-squared error in the Laplacian of the flux
function by the mean-squared Laplacian itself,
$\Lloss_{\rm Jz}=\langle(\Lap\hat\psi-\Lap\psi)^2\rangle/
\langle(\Lap\psi)^2\rangle$, and enters with weight
$w_{\rm Jz}=5\times10^{-2}$. Unlike the strong-form residuals, this term is
evaluated with the parity-aware spectral Laplacian rather than with central
differences, because the second derivative of $\psi$ across the thin layer is
the quantity the second-order stencil represents least well.

\section{Stacked residual and weighted-loss equivalence}\label{app:gn}

Dividing each equation's mean squared residual by the mean-squared magnitude of
its own additive terms, as in Eq.~\eqref{eq:phys-norm}, is a diagonal
reweighting of a nonlinear least-squares problem. Its effect on training is
clearest in a Gauss-Newton interpretation, which we set out here.

Evaluate the seven residuals $\R_i$ of Eq.~\eqref{eq:residuals} at every
interior collocation point and stack them into a single vector
$\br(\theta)\in\mathbb{R}^{N}$, ordered so that the entries belonging to one
equation occupy one contiguous block. The unnormalized physics loss is the
plain mean square $\tfrac1N\lVert\br\rVert^2$, while the normalized loss used
here is the weighted norm
\begin{equation}
  \Lloss_{\rm phys}
  =\tfrac1N\,\br^{\!\top}\bD\,\br,
  \qquad
  \bD=\operatorname{diag}(d_i^{-1}),
  \label{eq:weightednorm}
\end{equation}
where $\bD$ is constant within each equation block and
$d_i=\sum_k\langle(\R_i^{(k)})^2\rangle_{\Omega_{\rm crop}}+\epsilon$ is the
normalizing denominator of Eq.~\eqref{eq:phys-norm}. With
$\bJ=\pp\br/\pp\theta$ as the residual
Jacobian, the Gauss-Newton update that minimizes Eq.~\eqref{eq:weightednorm}
solves
\begin{equation}
  \big(\bJ^{\!\top}\bD\,\bJ\big)\,\delta\theta=-\,\bJ^{\!\top}\bD\,\br,
  \label{eq:gn}
\end{equation}
Thus, $\bD^{1/2}$ acts as a fixed left preconditioner on the residual, equivalent
to scaling each block of $\br$ by $d_i^{-1/2}$ before forming the normal
equations. This weighting does not change stationary points with
$\br=\bzero$; it rescales the curvature seen by the optimizer. Residual blocks
that differ by many orders of magnitude then contribute more evenly, allowing
the smaller induction residual to affect $\delta\theta$. We train with a
first-order method, so $\bD$ enters through the gradient
$\tfrac2N\bJ^{\!\top}\bD\,\br$. Equation~\eqref{eq:gn} gives the corresponding
second-order interpretation.

\section{Reproducibility and hyperparameters}\label{app:repro}

Table~\ref{tab:hyper} gives the complete configuration of the surrogate
reported in this paper, so that the model can be reproduced without reference
to the main text. The operator is a two-dimensional Fourier neural operator of
latent width $64$ and depth $4$, retaining $48$ Fourier modes per axis, with
reflection padding equal to $1/16$ of the grid on each side and a GELU
nonlinearity, giving $1.51\times10^{8}$ trainable parameters. It takes
$11$ input channels: the seven standardized initial-condition fields, one
constant channel carrying $\log_{10}S-4$, one constant channel carrying the
rescaled query time $2t/T-1$, and two coordinate channels $(x,y)$.
\par
Training uses a single fixed objective: a relative $L^2$ data term of unit
weight, a Sobolev ($H^1$) derivative term of weight $0.5$, a current-density
supervision term of weight $5\times10^{-2}$, an initial-condition
self-consistency term of weight $0.5$, and the normalized strong-form physics
residual of weight $10^{-2}$. The dataset comprises $34$ trajectories, each
with $41$ snapshots at $\Delta t=0.05$, computed on a $2049^2$ mesh and stored
at $513^2$; the two Lundquist numbers $S=1.1\times10^{4}$ and
$S=7.6\times10^{4}$ are withheld from training and used for evaluation.

\section{Notation}\label{app:notation}

Tables~\ref{tab:notation-phys} and \ref{tab:notation-op} shows every symbol
used in this paper: the first covers the physical fields, parameters, and
diagnostics, the second the neural-operator and training symbols. All
quantities are nondimensional in the units of Ref.~\cite{huang2016}, in which
the reference length, velocity, density, magnetic field, and pressure are all
set to unity ($L_*=V_*=\rho_*=B_*=p_*=1$) and time is measured in Alfv\'en
times.

\begin{table}[t]
\caption{\label{tab:notation-phys}Physical fields, parameters, and
diagnostics.}
\begin{ruledtabular}
\begin{tabular}{ll}
Symbol & Meaning \\
\hline
\multicolumn{2}{l}{\textit{Fields and derived quantities}}\\
$\rho$ & mass density \\
$p$ & thermal pressure \\
$\bv=(u,v,w)$ & velocity; in-plane $(u,v)$, out-of-plane $w$ \\
$\bm m=\rho\bv$ & momentum density \\
$\psi$ & in-plane magnetic flux function \\
$g\equiv B_z$ & out-of-plane (guide) magnetic field \\
$\bB=(B_x,B_y,g)$ & magnetic field, $B_x=-\pp_y\psi$, $B_y=\pp_x\psi$ \\
$J_z=\Lap\psi$ & out-of-plane current density \\
$P_T$ & total pressure, $p+\tfrac12(B_x^2+B_y^2+g^2)$ \\
$q$ & primitive state $(\rho,u,v,w,p,\psi,g)$ \\
\hline
\multicolumn{2}{l}{\textit{Parameters and scales}}\\
$\gamma=5/3$ & adiabatic index \\
$\eta$ & resistivity (magnetic diffusivity) \\
$\nu$ & kinematic viscosity \\
$\mathrm{Pr}_m=\nu/\eta$ & magnetic Prandtl number (unity) \\
$S=1/\eta$ & Lundquist number, $L_*V_A/\eta$ \\
$S_c\approx4\times10^{4}$ & critical $S$ for plasmoid onset \\
$V_A$ & Alfv\'en speed, $B_*/\sqrt{\rho_*}$ \\
$\tau_A$ & Alfv\'en time, $L_*/V_*$ \\
$c_s$ & sound speed, $\sqrt{\gamma p/\rho}$ \\
$c_f$ & fast magnetosonic speed, $\sqrt{c_s^2+|\bB|^2/\rho}$ \\
$\delta_{\rm SP}\sim S^{-1/2}$ & Sweet-Parker layer half-width \\
$h$ & initial current sheet half-thickness \\
$L_*,V_*,\rho_*,B_*,p_*$ & reference units (set to unity) \\
\hline
\multicolumn{2}{l}{\textit{Domain, operators, and diagnostics}}\\
$\Omega=[-\tfrac12,\tfrac12]^2$ & spatial domain \\
$\pp\Omega$ & domain boundary \\
$T$ & integration time \\
$\OmegaT=\Omega\times[0,T]$ & space-time domain \\
$\bn,\btau$ & wall normal, wall tangential unit vectors \\
$\pp_n$ & wall-normal derivative \\
$\grad$ & gradient \\
$\Lap=\pp_x^2+\pp_y^2$ & perpendicular Laplacian \\
$\Psi_{\rm rec}$ & reconnected flux, peak-to-peak $\psi$ on $|x|\le x_h$, $y=0$ \\
$\mathcal{R}_{\rm rec}$ & layer-averaged $\pp_t\psi$ on the same segment \\
$x_h=0.25$ & half-width of the $y=0$ diagnostic layer \\
\end{tabular}
\end{ruledtabular}
\end{table}

\begin{table}[t]
\caption{\label{tab:notation-op}Neural operator and training symbols.}
\begin{ruledtabular}
\begin{tabular}{ll}
Symbol & Meaning \\
\hline
\multicolumn{2}{l}{\textit{Operator and architecture}}\\
$\mathcal{G}_\theta$ & neural operator (parametric solution map) \\
$\theta$ & trainable parameters \\
$q_0=q(\cdot,0)$ & initial condition \\
$\hat q$ & predicted field \\
$n$ & grid points per side ($513$) \\
$\mathcal{F},\mathcal{F}^{-1}$ & forward, inverse discrete Fourier transform \\
$\mathcal{K}_\ell$ & spectral convolution at layer $\ell$ \\
$R_\ell$ & learnable spectral weights \\
$W_\ell$ & $1\times1$ (pointwise) convolution \\
$V^{(1)}_\ell,V^{(2)}_\ell$ & pointwise channel-mixing maps \\
$\sigma$ & activation function (GELU) \\
$w$ & latent width ($64$) \\
$L$ & number of Fourier layers ($4$) \\
$K_{\max}$ & retained Fourier modes per axis ($48$) \\
$\alpha$ & reflection-padding fraction ($1/16$) \\
\hline
\multicolumn{2}{l}{\textit{Training objective}}\\
$\Omega_{\rm int}$ & grid interior (boundary excluded) \\
$\Omega_{\rm crop}$ & cropped interior (wall strip removed) \\
$\Lloss$ & total training loss \\
$\Lloss_{L^2}$ & relative-$L^2$ data term \\
$\Lloss_{H^1}$ & Sobolev ($H^1$) derivative term \\
$\Lloss_{\rm Jz}$ & current-density supervision term \\
$\Lloss_{\rm ic}$ & initial-condition self-consistency term \\
$\Lloss_{\rm phys}$ & strong-form physics residual term \\
$\lambda_{H^1},w_{\rm Jz},w_{\rm ic},w_{\rm phys}$ & loss weights \\
$\R_i,\ \R_i^{(k)}$ & strong-form residual of equation $i$, and its $k$th additive term \\
$\epsilon$ & small stabilizing constant \\
\end{tabular}
\end{ruledtabular}
\end{table}

\begin{table}[t]
\caption{\label{tab:hyper}Configuration of the surrogate reported in
this paper. The single reported model is fully specified by these values.}
\begin{ruledtabular}
\begin{tabular}{ll}
Setting & Value \\
\hline
Grid $n$                      & $513$ \\
Retained modes $K_{\max}$     & $48$ \\
Latent width $w$              & $64$ \\
Fourier layers $L$            & $4$ \\
Channel-MLP expansion         & $1/2$ \\
Nonlinearity                  & GELU \\
Reflection padding $\alpha$   & $1/16$ \\
Input channels                & $11$ \\
Parameters                    & $1.51\times10^{8}$ \\
\hline
Data term weight              & $1$ (relative $L^2$) \\
$H^1$ weight $\lambda_{H^1}$  & $0.5$ \\
Physics weight $w_{\rm phys}$ & $10^{-2}$ (per-equation normalized) \\
Current weight $w_{\rm Jz}$     & $5\times10^{-2}$ \\
\hline
Optimiser                     & Adam, global-norm clip $1.0$ \\
Learning rate                 & $2\times10^{-3}$, cosine decay \\
Training steps                & $4\times10^{4}$ \\
Batch size                    & $6$ \\
Precision                     & single (\texttt{float32}) \\
Held-out $S$                  & $1.1\times10^{4},\ 7.6\times10^{4}$ \\
\hline
Trajectories                  & $34$ \\
Snapshots per trajectory      & $41$ ($\Delta t=0.05$) \\
DNS compute grid              & $2049^2$, saved at $513^2$ \\
\end{tabular}
\end{ruledtabular}
\end{table}

\bibliographystyle{apsrev4-2}
\bibliography{references}

\end{document}